{}

\documentclass[aps,prd,showpacs,superscriptaddress]{revtex4-1}  

\newif\ifpublic\publictrue
\newif\iffancy\fancytrue

\usepackage[a4paper,text={160mm,247mm},centering]{geometry}

\usepackage[T1]{fontenc}
\usepackage[utf8]{inputenc}

\usepackage{amsmath,amssymb, amsfonts, geometry}

\makeatletter
\providecommand*{\shuffle}{%
  \mathbin{\mathpalette\shuffle@{}}%
}
\newcommand*{\shuffle@}[2]{%
  \sbox0{$#1\vcenter{}$}%
  \kern .15\ht0 
  \rlap{\vrule height .25\ht0 depth 0pt width 2.5\ht0}%
  \raise.1\ht0\hbox to 2.5\ht0{%
    \vrule height 1.75\ht0 depth -.1\ht0 width .17\ht0 %
    \hfill
    \vrule height 1.75\ht0 depth -.1\ht0 width .17\ht0 %
    \hfill
    \vrule height 1.75\ht0 depth -.1\ht0 width .17\ht0 %
  }%
  \kern .15\ht0 
}
\makeatother

\usepackage{xparse}
\ExplSyntaxOn
\NewDocumentCommand{\Ef}{m m m}
 {
   \EE\!\left(\begin{smallmatrix}
  \Ef_print:n {#1} \\
  \Ef_print:n {#2}
\end{smallmatrix};#3\right)
 }
\seq_new:N \l_Ef_list_seq
\cs_new_protected:Npn \Ef_print:n #1
{
  \seq_set_split:Nnn \l_Ef_list_seq { , } { #1 }
  \seq_use:Nn \l_Ef_list_seq { & }
}
\ExplSyntaxOff

\ExplSyntaxOn
\NewDocumentCommand{\Efreg}{m m m}
 {
   \EEr\!\left(\begin{smallmatrix}
  \Efreg_print:n {#1} \\
  \Efreg_print:n {#2}
\end{smallmatrix};#3\right)
 }
\seq_new:N \l_Efreg_list_seq
\cs_new_protected:Npn \Efreg_print:n #1
{
  \seq_set_split:Nnn \l_Efreg_list_seq { , } { #1 }
  \seq_use:Nn \l_Efreg_list_seq { & }
}
\ExplSyntaxOff
\ExplSyntaxOn
\NewDocumentCommand{\Eval}{m m}
 {
   \EEv\left(\begin{smallmatrix}
  \Eval_print:n {#1} \\
  \Eval_print:n {#2}
\end{smallmatrix}\right)
 }
 \seq_new:N \l_Eval_list_seq
 \cs_new_protected:Npn \Eval_print:n #1
{
    \seq_set_split:Nnn \l_Eval_list_seq { , } { #1 }
    \seq_use:Nn \l_Eval_list_seq { , & }
}
\ExplSyntaxOff

\usepackage[pdfencoding=auto,bookmarks=true,hyperfigures=true]{hyperref}
\PassOptionsToPackage{unicode}{hyperref}
\usepackage{graphicx}
\usepackage{float}
\usepackage{lmodern}
\usepackage{amsbsy}
\usepackage{diagbox}
\usepackage{array}

\usepackage[usenames,dvipsnames]{xcolor}
\definecolor{dgreen}{rgb}{0,0.70,0.30}
\definecolor{gold}{rgb}{0.85,.66,0}
\definecolor{purple}{rgb}{1.0,0.3,0.6}

\makeatletter
\newsavebox{\apb@box}\newlength{\apb@width}
\newcommand{\autoparbox}[2][c]{\sbox{\apb@box}{#2}%
 \settowidth{\apb@width}{\usebox{\apb@box}}%
 \parbox[#1]{\apb@width}{\usebox{\apb@box}}}

\makeatother

\iffancy

\def\showkeysrefformat#1{{\normalfont\tiny\ttfamily#1}}
\makeatletter
\def\SK@@ref#1>#2\SK@{%
 {\@inlabelfalse\leavevmode\vbox to\z@{%
 \vss\SK@refcolor\rlap{\vrule\raise .75em%
  \hbox{\showkeysrefformat{#2}}}}}}
\makeatother
\fi

\numberwithin{equation}{section}

\newcommand{\eqn}[1]{eq.~\eqref{#1}}

\newcommand{\eqns}[2]{eqs.~\eqref{#1} and~\eqref{#2}}

\providecommand{\href}[2]{#2}

\makeatletter
\def\mr@ignsp#1 {\ifx\:#1\@empty\else #1\expandafter\mr@ignsp\fi}%
\newcommand{\multiref}[1]{\begingroup
\xdef\mr@no@sparg{\expandafter\mr@ignsp#1 \: }%
\def\mr@comma{}%
\@for\mr@refs:=\mr@no@sparg\do{\mr@comma\def\mr@comma{,}\ref{\mr@refs}}%
\endgroup}
\renewcommand{\eqref}[1]{(\multiref{#1})}
\makeatother

\makeatletter
\newcommand{\namedref}[2]{\hyperref[#2]{#1~\ref*{#2}}}
\newcommand{\secref}{\@ifstar{\namedref{Section}}{\namedref{section}}}
\newcommand{\subsecref}{\@ifstar{\namedref{Subsection}}{\namedref{subsection}}}
\newcommand{\appref}{\@ifstar{\namedref{Appendix}}{\namedref{appendix}}}
\newcommand{\tabref}{\@ifstar{\namedref{Table}}{\namedref{table}}}
\newcommand{\figref}{\@ifstar{\namedref{Figure}}{\namedref{figure}}}
\makeatother

\newcommand{\rcite}[1]{ref.~\cite{#1}}

\providecommand{\hypersetup}[1]{}

\hypersetup{plainpages=false}
\hypersetup{pdfpagemode=UseNone}
\hypersetup{bookmarksnumbered=true}
\hypersetup{pdfstartview=FitH}
\hypersetup{colorlinks=true, urlcolor=black, linkcolor=black, citecolor=[rgb]{0.15,0.35,0.65}}

\makeatletter
\let\@keywords\@empty
\let\@subject\@empty
\providecommand{\keywords}[1]{\gdef\@keywords{#1}}
\providecommand{\subject}[1]{\gdef\@subject{#1}}
\def\thetitle{\@title}
\def\theauthor{\@author}
\def\thesubject{\@subject}
\def\thedate{\@date}
\def\thekeywords{\@keywords}
\makeatother
\AtBeginDocument{
\hypersetup{pdftitle={\thetitle}}%
\hypersetup{pdfauthor={\theauthor}}%
\hypersetup{pdfsubject={\thesubject}}%
\hypersetup{pdfkeywords={\thekeywords}}%
}

\RequirePackage{verbatim}

\makeatletter
\newwrite\bibinl@out
{\immediate\closeout\bibinl@out\@esphack}
\makeatother

\newif\ifnote 
\notetrue

\allowdisplaybreaks

\usepackage{upgreek}

\def\Im{{\rm Im\,}}

\newcommand{\dimeps}{\epsilon}

\newcommand{\ve}{\varepsilon}

\newcommand{\dd}{\mathrm{d}}
\newcommand{\Dd}{\mathfrak{D}}

\newcommand{\CC}{\mathcal{C}}       %
\newcommand{\CF}{\mathcal{F}}       

\newcommand{\CI}{\mathcal{I}}

\newcommand{\CO}{\mathcal{O}}      

\newcommand{\CU}{\mathcal{U}}       

\newcommand{\CX}{\mathcal{X}}

\newcommand{\ZC}{\mathbb C}

\newcommand{\ZZ}{\mathbb Z}

\newcommand{\nnl}{\nonumber\\}

\RequirePackage{amsmath}
\makeatletter
\newcommand{\brk@ord}{\bBigg@{0}}
\newcommand{\brk@ordl}{\mathopen\brk@ord}
\newcommand{\brk@ordr}{\mathclose\brk@ord}
\newcommand{\brk@ordm}{\mathrel\brk@ord}
\newcommand{\brk@var}{\brk@ord}
\newcommand{\brk@varl}{\left}
\newcommand{\brk@varr}{\right}
\newcommand{\brk@varm}{\mathrel\brk@var}
\newcommand{\brk@altname}[3]{\expandafter\def\csname#2\expandafter\@gobble\string#1\endcsname{#1[#3]}}
\newcommand{\brk@usearg}[3]{%
  \def\brk@star{*}\def\brk@blank{}\def\brk@arg{#1}%
  \ifx\brk@arg\brk@blank\def\brk@arg{brk@ord}\fi%
  \ifx\brk@arg\brk@star\def\brk@arg{brk@var}\fi%
  \csname\brk@arg #2\endcsname#3}

\newcommand{\DeclareMathBrackets}[3]{
  \newcommand{#1}[2][]{\brk@usearg{##1}{l}{#2}##2\brk@usearg{##1}{r}{#3}}
  \brk@altname{#1}{big}{big}\brk@altname{#1}{lr}{*}}
\newcommand{\DeclareMathBiBrackets}[4]{
  \newcommand{#1}[3][]{\brk@usearg{##1}{l}{#2}##2#3##3\brk@usearg{##1}{r}{#4}}
  \brk@altname{#1}{big}{big}\brk@altname{#1}{lr}{*}}
\newcommand{\DeclareMathBiMBracketsStar}[4]{
  \newcommand{#1}[3][]{\brk@usearg{##1}{l}{#2}##2\brk@usearg{##1}{m}{#3}##3\brk@usearg{##1}{r}{#4}}
  \brk@altname{#1}{bi}{big}}
\newcommand{\DeclareMathBiBracketsStar}[4]{
  \newcommand{#1}[3][]{\brk@usearg{##1}{l}{#2}##2\brk@usearg{##1}{}{#3}##3\brk@usearg{##1}{r}{#4}}
  \brk@altname{#1}{big}{big}}

\makeatother

\DeclareMathBrackets{\brk}{(}{)}
\DeclareMathBrackets{\sbrk}{[}{]}
\DeclareMathBiBrackets{\comm}{[}{,}{]}
\DeclareMathBiBrackets{\acomm}{\{}{,}{\}}
\DeclareMathBiBrackets{\poiss}{\{}{,}{\}}
\DeclareMathBrackets{\abs}{|}{|}
\DeclareMathBrackets{\norm}{\lVert}{\rVert}
\DeclareMathBrackets{\eval}{.}{|}
\DeclareMathBrackets{\vev}{\langle}{\rangle}
\DeclareMathBrackets{\set}{\{}{\}}
\DeclareMathBiBrackets{\setcond}{\{}{;}{\}}
\DeclareMathBrackets{\bra}{\langle}{|}
\DeclareMathBrackets{\ket}{|}{\rangle}
\DeclareMathBiBracketsStar{\braket}{\langle}{|}{\rangle}
\DeclareMathBrackets{\floor}{\lfloor}{\rfloor}
\DeclareMathBrackets{\ceil}{\lceil}{\rceil}

\DeclareMathOperator{\EE}{E_4}
\DeclareMathOperator{\EEr}{E_4^\text{reg}}
\DeclareMathOperator{\EEv}{\ve_4}

\newcommand{\dx}{\,dx}

\newcommand{\dt}{\,dt}

\newcommand{\du}{\,du}

\usepackage{nicefrac}

\newcommand{\qand}{\quad\textrm{and}\quad}

\newcommand{\Sunrise}[1]{ 
\mbox{\parbox{2.5cm}{\hspace{0.25cm} 
\begin{picture}(2,1) 
\thicklines 
\put(0.0,0.5){\vector(1,0){0.1}} 
\put(-0.5,0.5){\line(1,0){3.0}} 
\put(1,0.5){\circle{1.6}} 
\put(0.82,1.4){$m_1$} 
\put(0.82,0.62){$m_2$} 
\put(0.82,-0.15){$m_3$} 
\put(0.0,0.7){\makebox(0,0)[b]{$#1$}} 
\end{picture} 
}} 
\hfill} 

\newcommand{\ifequals}[3]{\ifthenelse{\equal{#1}{#2}}{#3}{}}
\newenvironment{switch}[1]{\renewcommand{\case}{\ifequals{#1}}}{}

\newcommand{\me}[2]{
    \begin{switch}{#1}
        \case{1}{\Ef{-2}{0}{#2}}
        \case{2}{\Ef{-2}{1}{#2}}
        \case{3}{\Ef{-2}{\infty}{#2}}
        \case{4}{\Ef{-1}{0}{#2}}
        \case{5}{\Ef{-1}{1}{#2}}
        \case{6}{\Ef{-1}{\infty}{#2}}
        \case{7}{\Ef{0}{0}{#2}}
        \case{8}{\Ef{1}{1}{#2}}
        \case{9}{\Ef{1}{\infty}{#2}}
        \case{10}{\Ef{0,-1}{0,0}{#2}}
        \case{11}{\Ef{0,-1}{0,1}{#2}}
        \case{12}{\Ef{0,-1}{0,\infty}{#2}}
        \case{13}{\Ef{0,0}{0,0}{#2}}
        \case{14}{\Ef{1,0}{0,0}{#2}}
    \end{switch}
}

\begin{document}
\setlength{\unitlength}{1.0cm} 
\hspace{1cm}
  CERN-2017-274\quad
  CP3-17-58\quad
HU-EP-17/30\quad
HU-Mathematik-2017-10\quad
  SLAC-PUB-17195


\title{\textbf{Elliptic polylogarithms and iterated integrals on elliptic curves II: \\an application to the sunrise integral}}

\author{Johannes Broedel}
\email{jbroedel@physik.hu-berlin.de}
\affiliation{
Institut f\"ur Mathematik und Institut f\"ur Physik,
Humboldt-Universit\"at zu Berlin,\\
IRIS Adlershof, Zum Gro\ss{}en Windkanal 6, 12489 Berlin, Germany}

\author{Claude Duhr}
\email{claude.duhr@cern.ch}
\affiliation{Theoretical Physics Department,
CERN, CH-1211 Geneva 23, Switzerland.
}
\affiliation{Center for Cosmology, Particle Physics and Phenomenology (CP3),\\
Universit\'e catholique de Louvain,\\
Chemin du Cyclotron 2, 1348 Louvain-La-Neuve, Belgium.
}

\author{Falko Dulat}
\email{dulatf@slac.stanford.edu}
\affiliation{
SLAC National Accelerator Laboratory, Stanford University,
Stanford, CA-94309, USA}

\author{Lorenzo Tancredi}
\email{lorenzo.tancredi@cern.ch}
\affiliation{Theoretical Physics Department,
CERN, CH-1211 Geneva 23, Switzerland.
}
\vskip1cm

\begin{abstract}
\vskip2cm
We introduce a class of iterated integrals that generalize multiple polylogarithms to
elliptic curves.
These elliptic multiple polylogarithms are closely related to similar functions defined in pure mathematics and string theory.
We then focus on the equal-mass and non-equal-mass sunrise integrals, and we develop
a formalism that enables us to compute these Feynman integrals in terms of
our iterated integrals on elliptic curves.
The key idea is to use integration-by-parts identities to identify a set of integral
kernels, whose precise form is determined by the branch points of the integral in question. These kernels
allow us to express all iterated integrals on an elliptic curve in terms of them.
The flexibility of our approach leads us to expect that it will be applicable to a large variety of
integrals in high-energy physics.
\end{abstract}

\vspace*{2cm}

\maketitle

\setcounter{tocdepth}{2}
\tableofcontents
\hypersetup{linkcolor=[rgb]{0.15,0.35,0.65}}

\section{Introduction}
\label{sec:introduction}

With the discovery of the Higgs boson at the Large Hadron Collider (LHC) at CERN and the absence of signals of physics beyond the Standard Model (SM),
 we have entered a new era of precision physics. Indeed, the mass of Higgs boson was the last free parameter of the SM
 whose value had not yet been determined by experiments. Therefore, with the measurement of the Higgs mass, the SM 
 is a fully predictive theory that can be valid, at least in principle, all the way up to the Planck scale. Even if the scale of new physics 
 is beyond the reach of the LHC, small deviations to SM predictions may
 still show up in total cross sections and distributions, due to the effect of virtual particles in the loops. It is therefore of paramount importance 
 that we are able to provide very precise theoretical predictions which match the precision reached by experimental measurements at the LHC.

Precision calculations for LHC observables require the evaluation of Feynman diagrams where
additional virtual or real particles are present in the process. While next-to-leading order computations, which 
include contributions with one additional real or virtual particle, have been fully automated a few years ago already,
we are currently living through an era where next-to-next-to-leading order computations are becoming the new standard. 
Key in this progress has been, among other things, an improved understanding of how to perform two-loop computations.
In particular, it has become clear that large classes of Feynman integrals can be expressed in terms of a class of special
functions known as \emph{multiple polylogarithms} (MPLs)~\cite{Goncharov:1995,Remiddi:1999ew,Goncharov:2001}. One of the main advantages
of working with MPLs is that their underlying mathematical and algebraic structures are well understood~\cite{Goncharov:2010jf,Ablinger:2011te,Goncharov:2010jf,Duhr:2011zq,Duhr:2012fh,Duhr:2014woa}.

It is well known, however, that not all two-loop Feynman integrals can be expressed in terms of MPLs alone. 
The first appearance of a two-loop integral that cannot be expressed in terms of MPLs goes back to QED~\cite{Sabry}, where it was observed that the electron self-energy
involves integrals of elliptic type. 
Similar functions have recently shown up in two-loop computations relevant to LHC processes, like for example
top-quark pair production~\cite{Czakon:2008ii,vonManteuffel:2017hms,Hidding:2017jkk}, Higgs production in association with a jet~\cite{Bonciani:2016qxi}
and top-mass effects in diphoton and dijet production~\cite{Becchetti:2017abb}. The previous processes all explicitly involve a heavy massive particle in the loop, and it was assumed for a long time that massive propagators are a necessary condition for the appearance of non-polylogarithmic structures. Recently, however, it was shown that
elliptic integrals also show up in planar $\mathcal{N}=4$ Super Yang-Mills~\cite{CaronHuot:2012ab,Nandan:2013ip,Bourjaily:2017bsb}. 
Although the previous examples make it clear that progress in multi-loop computations will require a deeper understanding of Feynman integrals of elliptic type, still very little is known about the mathematical properties of the resulting class of functions. This, in turn, has sparked a lot of activity in recent years in trying to uncover the mathematics of elliptic Feynman integrals~\cite{Broadhurst:1987ei,Bauberger:1994by,Bauberger:1994hx,Laporta:2004rb,Kniehl:2005bc,Bloch:2013tra,Adams:2013nia,Remiddi:2013joa,Adams:2014vja,Adams:2015gva,Adams:2015ydq,Remiddi:2016gno,Adams:2016xah,Remiddi:2017har,Hidding:2017jkk,Passarino:2016zcd,Chen:2017soz}

The most prominent elliptic Feynman integral is the so-called sunrise integral, i.e., the two-loop integral with three massive propagators. 
It had been observed already fifteen years ago that the maximal cut of this integral can be expressed in terms of complete
elliptic integrals of the first kind~\cite{Laporta:2004rb}. The result for the full (uncut) integral, however, remained mysterious for more than a decade. 
In a landmark paper Bloch and Vanhove have shown that the sunrise integral with three equal masses in two dimensions can naturally be written 
in terms of a generalization of the dilogarithm to an elliptic curve~\cite{Bloch:2013tra}. The latter is a special case of a more general class of functions, called \emph{elliptic multiple polylogarithms (eMPLs)}~\cite{BeilinsonLevin,LevinRacinet,BrownLevin}, and they have recently appeared also in the context of superstring amplitudes at one-loop~\cite{Broedel:2014vla,Broedel:2015hia,Broedel:2017jdo}. The result of ref.~\cite{Bloch:2013tra} has sparked a wealth of new results and representations for the sunrise integral, including 
also higher-order terms in dimensional regularization and results in four space-time dimensions~\cite{Adams:2013nia,Remiddi:2013joa,Adams:2014vja,Adams:2015gva,Adams:2015ydq,Remiddi:2016gno,Adams:2016xah,Remiddi:2017har,Hidding:2017jkk}. A common feature of these results is that most of them require the introduction of a new elliptic generalization of MPLs, whose relationship to the eMPLs that have appeared in pure mathematics and string theory is often unclear.
This is somewhat disconcerting, because in the non-elliptic case it was precisely the realisation that ordinary MPLs constitute the right class of functions with beautiful algebraic properties that was at the heart of a lot of progress in multi-loop computations.

In the present paper, we try to close this gap, and we introduce a class of functions that are defined as iterated integrals on an elliptic curve. The ensuing functions have at most logarithmic singularities -- thereby constituting a genuine generalization of polylogarithms to elliptic curves. We discuss how one can easily compute the sunrise integral in term of these functions, and we present analytic results for all the master integrals of the sunrise topology in $d=2-2\dimeps$ dimensions. In particular, we present for the first time an analytic expression for the second master integral in the case of three unequal masses. In a companion paper~\cite{longelliptic}, we study in detail some of the properties of our functions. In particular, we show that they are equivalent to the eMPLs introduced in the mathematics literature. As such, our functions genuinely deserve being called elliptic multiple polylogarithms as well. At the same time, this shows how the sunrise integral is connected to the eMPLs that have appeared in mathematics and string theory.

The outline of the paper is as follows: after providing a lightning overview
of some background on the sunrise
integrals in \secref{sec:sunrise}, we will jump into the evaluation of the
first master integral for the equal-mass sunrise integral in \secref{sec:firstmaster}.
This integral will serve as our prime example of how eMPLs naturally arise in the context of 
the sunrise integrals.
After this first
encounter with iterated integrals on elliptic curves, we will discuss and compute the
second master integral for the equal-mass sunrise integral in \secref{sec:secondmaster}.
We will collect structural results of the first sections, including the
complete set of integration kernels that define eMPLs, in a brief summary section \ref{sec:recap}. In
\secref{sec:application}, we will apply our new language to the more complex
scenario of sunrise integrals with three different masses. In particular, we will discuss the unitary cut of the sunrise integral as well as the
unequal-mass master integrals for the sunrise topology from dispersion relations. In \secref{sec:conclusion} we draw our conclusion. 

\section{The sunrise integral: overview}
\label{sec:sunrise}

The most popular example of a family of Feynman integrals that cannot be computed in terms
of multiple polylogarithms are the \textit{sunrise integrals}, which have received a lot of attention over the last few years. 
The sunrise integrals can be represented by the following graph,
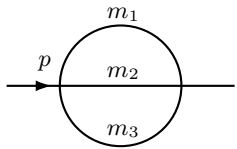
\begin{figure}[!h]
\label{fig:sunrise}
$$ \Sunrise{p} $$ 
\end{figure}

\noindent and the corresponding family of Feynman integrals reads
\begin{equation}
  \label{eqn:sunrisegeneral}
  S_{\nu_1\nu_2\nu_3}(S,m_1^2,m_2^2,m_3^2)= \int \frac{\Dd^dk_1\,\Dd^dk_2}{(k_1^2-m_1^2)^{\nu_1}(k_2^2-m_2^2)^{\nu_2}((k_1-k_2+p)^2-m_3^2)^{\nu_3}}\,,
\end{equation}
where the integration measure is defined as
\begin{equation}
\int \Dd^d k \equiv e^{\gamma_E\dimeps}\int \frac{\dd^d k}{i\, \pi^{d/2}}\,,
\end{equation}
$\gamma_E=-\Gamma'(1)$ is the Euler-Mascheroni constant
and $\nu_i\in\ZZ$ denote the multiplicities of the propagators. We work in dimensional
regularization in $d=d_0-2\dimeps$ dimensions, where $d_0$ is even. 
We define $S=-s=-p^2$, and the quantities
$k_i$ and $m_i$ denote the loop momenta and the masses of the propagators
respectively. 

In terms of Feynman parameters, the integral in~\eqn{eqn:sunrisegeneral} can
equivalently be written as
\begin{equation}\begin{split}
  \label{eqn:sunriseFeynman}
  \!\!S_{\nu_1\nu_2\nu_3}&(S,m_1^2,m_2^2,m_3^2)  =\frac{e^{2\gamma_E\dimeps}\Gamma(\nu-d)(-1)^{\nu}}{\Gamma(\nu_1)\Gamma(\nu_2)\Gamma(\nu_3)}\int_0^{\infty}
\!\!\!  \dx_1\dx_2\dx_3\,
  x_1^{\nu_1-1}x_2^{\nu_2-1} x_3^{\nu_3-1} 
  \frac{\mathcal{U}^{\nu-3/2 d}}{\mathcal{F}^{\nu-d}}
  \delta(1-x_3)\,,
\end{split}
\end{equation}
with $\nu\equiv \sum_{i=1}^3\nu_i$, and the Symanzik polynomials are
\begin{equation}
  \label{eqn:Symanzik}
  \CU=x_1 x_2 + x_2 x_3 + x_1 x_3 \qand 
  \CF= x_1 x_2 x_3 S +(m_1^2 x_1 +m_2^2 x_2 + m_3^2 x_3)\,\CU\,.
\end{equation}

In order to get a feeling for the class of functions that show up when
evaluating the integral in~\eqn{eqn:sunriseFeynman}, let us first consider the case where all masses are
equal, $m_i = m\neq 0$. We will return to the unequal-mass case in
\secref{sec:application}. Throughout this paper, we only discuss the sunrise integrals in $d=2-2\dimeps$ dimensions.
This allows us to focus on the
elliptic core of the integrals. We stress that the restriction to
two dimensions is not a limitation, because the results in
$d=4-2\dimeps$ dimensions can always be recovered via dimensional recurrence
relations~\cite{Tarasov:1996br,Lee:2009dh}. 

Equation~\eqref{eqn:sunrisegeneral} defines an infinite family of Feynman
integrals,
but not all members of the family are independent.  We can use
integration-by-parts (IBP) identities to reduce any member in the sunrise
topology with three equal (non-zero) masses to a linear combination of three
master integrals, which we can choose to be $S_{111}$, $S_{211}$ and $S_{110}$.
$S_{110}$ is a product of one-loop integrals and will not be discussed any
further.  Note that, of course, in order to obtain an
integral family for the sunrise graph closed under IBP identities we 
need to add two independent
scalar products to the definition in~\eqn{eqn:sunrisegeneral}. 
In the following we will often refer to $S_{111}$ and $S_{211}$
loosely as the \emph{first} and \emph{second} master integrals.  These two
master integrals will be our guiding examples for exploring the world of
elliptic multiple polylogarithms in this article.  We stress that our paper is
not the first to consider the sunrise integrals, but they have been computed in
various different guises in terms of functions that cannot be expressed in
terms of multiple polylogarithms (MPLs). The goal of this paper is to
show that there is a natural class of iterated integrals on an elliptic curve through
which our results can be expressed. This class of iterated integrals, whose
mathematical properties are spelled out in more detail in a companion
paper~\cite{longelliptic}, have at most logarithmic singularities on the
elliptic curve, and as such they deserve to be called \emph{elliptic
multiple polylogarithms (eMPLs)}. Moreover, in ref.~\cite{longelliptic} we show that these
elliptic polylogarithms are indeed very closely related to the multiple
elliptic polylogarithms considered in the mathematics
literature~\cite{BeilinsonLevin,LevinRacinet,BrownLevin}.

In the next two sections we will describe in detail the computation of the first
and second master integrals.

\section{First master integral of the equal-mass sunrise}
\label{sec:firstmaster}

\subsection{The first master integral in $d=2$ dimensions}

Starting from \eqn{eqn:sunriseFeynman}, the first master integral of the equal-mass
sunrise topology admits the integral representation
\begin{equation}
  \label{eqn:sunriseequalmass}
  S_{111}(S,m^2) \equiv S_{111}(S,m^2,m^2,m^2) =- 
  \Gamma(3-d)e^{2\gamma_E\dimeps}\int_0^{\infty}\dx_1\dx_2\dx_3\,\delta(1-x_3)\frac{\mathcal{U}^{-3/2(d-2)}}{\mathcal{F}^{3-d}}\,.
\end{equation}
The integral is convergent in two dimensions, and so we can simply perform the
expansion in the dimensional regulator $\dimeps$ at the integrand level.  At
leading order in $\dimeps$, the integral $S_{111}$ is determined solely by the
polynomial $\CF$ defined in \eqn{eqn:Symanzik}: 
\begin{equation}
    \label{eq:SunriseEp0}
    S_{111}(S,m^2)\Bigr|_{\dimeps^0}
=-\int_0^{\infty}\dx_1\dx_2\frac{1}{m^2(x_1+x_2+1)s_2(x_1,x_2,1)+S\,x_1x_2}\,.
\end{equation}
The denominator is a quadratic polynomial in $x_2$. We perform the change of variables
\mbox{$x_1=x/(1-x)$}, and we find
\begin{equation}
    S_{111}(S,m^2)\Bigr|_{\dimeps^0}=-
    \int_0^1\!\!\dx\int_0^{\infty}\!\!\dx_2\,
    \frac{1}{m^2(x+x_2)(1+(1-x)x_2)+S(1-x)x\,x_2}\,.
\end{equation}
After linearizing the denominator by partial fractioning in $x_2$, the remaining integral
is easily performed, yielding
\begin{equation}
\label{eqn:ellipticIntegral}
S_{111}(S,m^2)\Bigr|_{\dimeps^0}
= \frac{1}{(m^2+S)}\int_0^1\!\!\dx\,\frac{\log\chi}{y}\,,
\end{equation}
with
\begin{align}
  \label{eqn:quantities}
  \chi &= \frac{x(x-1)(m^2+S)-m^2-(m^2+S)y}{x(x-1)(m^2+S)-m^2+(m^2+S)y}\,,
  \end{align}
  and
  \begin{align}
  y 
  &=\sqrt{(x-a_1)(x-a_2)(x-a_3)(x-a_4)}\,,
  \label{eqn:y}
\end{align}
where $a_i$, $1\le i\le 4$, denote the roots of the quartic polynomial inside
the square root
\begin{equation}\begin{split}
  \label{eqn:roots}
a_1 = \frac{1}{2}(1-\sqrt{1+\rho})\,,\quad & a_2 = \frac{1}{2}(1+\sqrt{1+\rho})\,,\quad a_3= \frac{1}{2}(1-\sqrt{1+\overline{\rho}})\,,\quad  a_4=\frac{1}{2}(1+\sqrt{1+\overline{\rho}})\,.
\end{split}\end{equation}
The auxiliary variables $\rho$ and $\bar{\rho}$ are defined by
\begin{equation}
    \rho = -\frac{4m^2}{(m+\sqrt{-S})^2}\qand \overline{\rho} = -\frac{4m^2}{(m-\sqrt{-S})^2}\,. 
\end{equation}
For real values of $\sqrt{-S}$ and $m$ they are complex conjugates of each
other.  The four roots $a_i$ of the quartic polynomial are clearly distinct for
distinct and non-vanishing values of $\sqrt{-S}$ and $m$, and therefore the quartic
polynomial is irreducible. 

In absence of the logarithm in the numerator in~\eqn{eqn:ellipticIntegral}, the integral would
evaluate to an \emph{incomplete elliptic integral of the first kind}:
\begin{equation}
  \label{eqn:ellK}
  \text{F}(x|w^2)=\int_0^x\frac{dt}{\sqrt{(1-t^2)(1-w^2 t^2)}}\,.
\end{equation}
While elliptic integrals of this type are well studied in 19th century mathematics, 
integrals of the type in~\eqn{eqn:ellipticIntegral}, with a logarithm in the
numerator, are not part of the classical literature on elliptic integrals;  only
recently a subclass of these integrals has been studied in detail in ref.~\cite{Remiddi:2017har}.
One of the main goals of this paper
is to show how such integrals can be
performed in complete generality in terms of a well-defined class of iterated integrals. 

We start by analyzing the logarithmic term in \eqn{eqn:ellipticIntegral}
separately.  Differentiating with respect to $x$ and integrating back, we find the following
integral representation for the logarithm,
\begin{equation}
    \label{eqn:logIntegrand}
    \log\chi
    = \int_0^{x}\frac{\dx'}{y_{x'}}\Biggl[2x'-1-\frac{m^2}{(m^2+S)(1-x')}+\frac{m^2}{(m^2+S)x'}\Biggr]\,.
    \end{equation}
Perhaps surprisingly, the logarithm itself can be written as an integral of
elliptic type, as indicated by $y_{x'}$ in the denominator (which denotes $y$
evaluated at $x'$). Concretely, \eqn{eqn:logIntegrand} contains three integrals of
elliptic type:
\begin{equation}
    \label{eqn:threeIntegrals}
    \int\frac{\dx}{y}\,,\qquad\int\frac{\dx}{y}x\,,\qquad\int\frac{\dx}{y}\frac{1}{x-c}\,.
\end{equation}
The first integral in~\eqn{eqn:threeIntegrals} bears some similarities with the
elliptic integral of the first kind K in~\eqn{eqn:ellK}. One is tempted
to try to integrate the remaining two integrals by parts in order to reduce
them to those of the first kind. That attempt is futile, however, and one can show that
indeed the three integrals in~\eqn{eqn:threeIntegrals} are independent
with respect to IBP identities. This, in turn, is a well-known result in mathematics: the second
and third integral in~\eqn{eqn:threeIntegrals} are related to the
\emph{incomplete elliptic integrals of third kind}:
\begin{equation}
  \label{eqn:ellPi}
  \Pi(n^2,x|w^2)=\int_0^x\frac{dt}{\sqrt{(1-t^2)(1-w^2 t^2)}}\frac{1}{1-n^2 t^2}\,.
\end{equation}
As a conclusion, all the integrations in~\eqn{eqn:logIntegrand} can be performed in terms of elliptic
integrals of the first and third kind, leading to an intriguing relation between a logarithm and a sum of elliptic integrals.

When we rewrite the logarithm in~\eqn{eqn:ellipticIntegral} as a combination of elliptic integrals,
the classical theory of elliptic integrals of the 19th century is no longer sufficient to perform the integral. 
Very loosely speaking, our goal is to extend
the theory of elliptic integrals to include those iterated integrations. This is achieved by defining an elliptic generalization of multiple
polylogarithms, \emph{elliptic multiple polylogarithms} (eMPLs), which are
defined as iterated integrals over suitable integration kernels
$\psi_i$,
\begin{equation}
  \label{eqn:ellPoly}
  \Ef{n_1,\ldots,n_k}{c_1,\ldots,c_k}{x}=\int_0^x\dt\,\,\psi_{n_1}(c_1,t) \Ef{n_2,\ldots,n_k}{c_2,\ldots,c_k}{t}\,,
\end{equation}
where $n_i\in\ZZ$, $c_i\in \ZC\cup\lbrace\infty\rbrace$ and the recursion
starts with $\EE(;x)=1$.  Note that the definition in~\eqn{eqn:ellPoly} is in
complete analogy with the ordinary multiple polylogarithms, which are defined
recursively by
\begin{equation}\label{eq:MPL_def}
    G(a_1,\dots,a_n; x) = \int_0^x\frac{\dx'}{x'-a_1}G(a_2,\dots,a_n;x')\,,
\end{equation}
starting with $G(;x) = 1$.

Let us briefly discuss some of the properties of the integration kernels $\psi_i$ in
\eqn{eqn:ellPoly}. The kernels should satisfy three basic requirements. First,
they should not be total derivatives, because otherwise the integration would
be trivial.  Second, to form a good basis, they should be independent with
respect to IBP identities. Finally, since our eMPLs should have at
most logarithmic singularities, the integration kernels should have at most
simple poles.  From the previous discussion, it follows that we have natural
candidates in~\eqn{eqn:threeIntegrals}  that satisfy all of the requirements.
We therefore define the following integration kernels,
\begin{equation}
    \label{eqn:firstkernels}
    \psi_0(0,x) = \frac{c_4}{y}\,,\quad\psi_{-1}(\infty,x)=\frac{x}{y}\,,\quad
    \psi_{-1}(c,x) = \frac{y_c}{(x-c)y}-\frac{\delta_{c0}}{x},
\end{equation}
where the kernel $\psi_0(0,x)$ includes the normalization factor $c_4 =
\frac{1}{2}\sqrt{a_{13}a_{24}}$, with $a_{ij}=a_i-a_j$.  Since our iterated integrals shall be a
generalization of ordinary MPLs, we include their integration kernel from
\eqn{eq:MPL_def}, and we define
\begin{equation}\label{eqn:psi1}
\psi_{1}(c,x)=\frac{1}{x-c}\,.
\end{equation}
Note that the previous equation implies that ordinary MPLs are just a subset of
eMPLs.  While these definitions may seem ad hoc at this point, we will argue in
more detail in \secref{sec:secondmaster} that these kernels indeed define
genuine eMPLs. In particular, we will see that we need to extend the set of
kernels in \eqn{eqn:firstkernels} in order to obtain a complete and independent
set of eMPLs. The kernels in \eqn{eqn:firstkernels}, however, are sufficient
for the computation of the first master integral of the equal-mass sunrise topology, and so we
defer a more detailed discussion of the complete set of integration kernels to a
subsequent section.

Let us also comment on the singularity structure of
the integration kernels $\psi_i$.  Different types of kernels are
indexed by an integer subscript, and the first argument of $\psi_i(c,x)$ indicates that this kernel has a simple pole at $x=c$. 
In particular, $c$ may be infinite, indicating that there is a
pole at $x=\infty$. Indeed, letting $x=1/u$,
we see that $\psi_{-1}(\infty,x)$ has a pole at $x=\infty$
(that shows up as a pole at $u=0$). The only kernel deviating from this
nomenclature is $\psi_0(0,x)$, which does not have any pole. Further
clarification is required for the kernel $\psi_{-1}(0,x)$: it is
straightforward to check that $1/(xy)$ has a pole $x=0$ (assuming $a_i\neq0$),
leading to an end-point singularity in \eqn{eqn:ellPoly} whenever $c_k=0$.  Correspondingly, the
pole at $x=0$ is removed by the Kronecker $\delta$ term, making the convergence of the integral manifest.  Note that also $\psi_{1}(0,x) = 1/x$ is
singular at $x=0$.  In this case, however, we do not subtract the singularity,
but we use the following special definition, well-known from the case of
ordinary MPLs,
\begin{equation}
    \label{eqn:logdef}
    \textrm{E}_4(\underbrace{\begin{smallmatrix}1\dots1\\0\dots0\end{smallmatrix}}_{n\textrm{ times}};x) = G(\underbrace{0\dots0}_{n\textrm{ times}};x)
    = \frac{1}{n!}\log^nx\,,\quad \log x=\int_1^{x}\frac{\dx'}{x}.
\end{equation}
Finally, by using $y$ in \eqn{eqn:firstkernels} we are implicitly using
information about the geometry of the underlying elliptic curve we are working on. Thus it would be
accurate to attach the root vector $\vec{a}=(a_1,a_2,a_3,a_4)$ as an additional
parameter to each of the kernels and E$_4$.   However, in order to avoid cluttering the
notation, we always suppress the dependence on the root vector, as in this paper 
all our equations 
will involve
objects defined on the same elliptic curve only (we assume the external kinematics fixed).

Armed with these definitions, we can now return to~\eqn{eqn:logIntegrand} and
write it in terms of the integration kernels,
\begin{equation}
    \log\chi
    = \int_0^{x}\dx'\,\Bigl[\frac{1}{c_4}\psi_0(0,x') -2
    \psi_{-1}(\infty,x')+\psi_{-1}(0,x')+\psi_{1}(0,x')+\psi_{-1}(1,x')\Bigr].
\end{equation}
Using the iterated definition of the eMPLs
in~\eqn{eqn:ellPoly}, all integrations can be performed with ease and yield
\begin{equation}
  \log\chi
  =\frac{1}{c_4}\Ef{0}{0}{x}-2\Ef{-1}{\infty}{x}+\Ef{-1}{0}{x}+\Ef{1}{0}{x}+\Ef{-1}{1}{x}.
  \label{eqn:logRewritten}
\end{equation}
The previous equation is a remarkable identity that allows us to write a logarithm 
with a complicated algebraic argument in terms of elliptic polylogarithms with simple arguments.
Inserting~\eqn{eqn:logRewritten} into~\eqn{eqn:ellipticIntegral}, we obtain
\begin{equation}\begin{split}
    S_{111}(S,m^2)\Bigr|_{\dimeps^0}=\frac{1}{(m^2+S)
        c_4}\int_0^1&\dx\,\psi_{0}(0,x)
    \Biggl[\frac{1}{c_4}\Ef{0}{0}{x}-2\Ef{-1}{\infty}{x}\\
    &+\Ef{-1}{0}{x}+\Ef{1}{0}{x}+\Ef{-1}{1}{x}\Biggr]\,,
    \end{split}
\end{equation}
where we have replaced ${1}/{y}$ with the integral kernel
$\psi_0(0,x)/c_4$. The remaining integrations can easily be performed using
\eqn{eqn:ellPoly}, which leads to the result
\begin{equation}\begin{split}
\label{eqn:S111_result}
S_{111}(S,m^2)\Bigr|_{\dimeps^0}=\frac{1}{(m^2+S)c_4}\Biggr[&\frac{1}{c_4}\Ef{0,0}{0,0}{1}-2\Ef{0,-1}{0,\infty}{1}-\Ef{0,-1}{0,0}{1}\\
&-\Ef{0,-1}{0,1}{1}-\Ef{0,1}{0,0}{1}\Biggr]\,.
    \end{split}
\end{equation}
Equation~\eqn{eqn:S111_result} is the final result for the first master integral of the sunrise 
topology in $d=2$ dimensions. As promised, we see that the result can be cast entirely in the form
of a linear combination of eMPLs defined in \eqn{eqn:ellPoly}.
Before we continue and discuss how to extend our framework to higher orders in
$\dimeps$ and the second master integral, let us summarize our main steps: we
were able to write a simple function, the logarithm appearing
in~\eqn{eqn:ellipticIntegral}, with complicated arguments involving a square
root of a quartic polynomial, in terms of more complicated functions, the
elliptic multiple polylogarithms, with a simple argument. This strategy is
reminiscent of the strategy employed successfully in the computation of
polylogarithmic integrals. A key ingredient in establishing our method was the
identification of the integration kernels $\psi_i$. The integral
considered so far was simple enough to appear directly in terms of these
kernels. However, as will be discussed below, it is possible to tackle more
complicated integrals using IBP identities. 


\subsection{The first master integral in dimensional regularization}

So far we have only discussed the master integral $S_{111}$ in strictly two dimensions.
In the remainder of this section we extend our analysis to higher orders in the $\dimeps$
expansion, and we show that $S_{111}$ can be expressed in terms of eMPLs at every order in
dimensional regularization. We only discuss the linear term in $\dimeps$ explicitly. The
extension to higher orders is straightforward.

Expanding the Feynman parameter integral in \eqn{eqn:sunriseequalmass}
to linear order in $\dimeps$, we find,
\begin{equation}
\label{eqn:S111_e1}
    S_{111}(S,m^2)\Bigr|_{\dimeps^1}
    = \int_0^{\infty}\dx_1\dx_2\dx_3\frac{\delta(1-x_3)[2\log(\CF)-3\log(\CU)]}{\CF}\,.
\end{equation}
The denominator in \eqn{eqn:S111_e1} is identical to the denominator in \eqn{eq:SunriseEp0},
and so we can expect that the integration can be done using a similar algorithm as for $\dimeps=0$. 
Hence, we let $x_1=x/(1-x)$, and the result of the
integral over $x_2$ can be written in terms of ordinary multiple
polylogarithms as 
\begin{align}
    \label{eqn:oepintegrand}
    S_{111}&(S,m^2)\Bigr|_{\dimeps^1}\!\!\!\!\!=\!\frac{1}{m^2+S}\!\int_0^1\!\!\frac{\dx}{y}\Biggl[G(0;\chi)\Bigl(
    G(0;\chi')\!
    +\!G(1;x)\!
    +\!4G(1;\chi)\!
    +\!2G(0;m^2)
    -\!3G\Bigl(\frac{\chi'}{\chi};x\Bigr)\!
    \Bigr)\\
  \nonumber  &\,
    -3G(0,0;\chi)\!
    +\!3G(0;\chi')G\Bigl(\frac{\chi'}{\chi};x\Bigr)\!\!
    -\!3G\Bigl(\frac{\chi'}{\chi},0;x\Bigr)\!\!
    -\!4G(0,1;\chi)
    -\!G(0;\chi')G(\chi';x)\!
    +\!3G(\chi',0;x)\!
    -\!4\zeta_2
    \Biggr]\,,
\end{align}
where $\chi$ has been defined in~\eqn{eqn:quantities}, and we have introduced the quantity
\begin{equation}
    \chi' = \frac{(m^2+S)y+m^2(x^2-x-1)+S(x-1)x}{2m^2(x-1)}\,.
\end{equation}
The remaining integration over $x$ can be done following the same steps as for $\dimeps=0$:
we write every ordinary MPL with a complicated argument in the integrand as a linear combination of eMPLs
such that the integration can immediately be performed using the definition of the E$_4$ functions in~\eqn{eqn:ellPoly}.
This rewriting can be done recursively in the weight of the MPLs. 
For example, let us consider how to rewrite $G(1;\chi)$ in terms of eMPLs. We can differentiate with
respect to $x$ and integrate back, and we find,
\begin{align} 
\nonumber  G&(1;\chi)
    = -i\pi+\int_0^x\frac{\dx'}{y_{x'}}\frac{(1-2x')(m^2(1-(1-x')x')-S(1-x')x')}{2(1-x')x'(m^2+S)} -\int_0^x\dx'\frac{1-2x'}{2x'(1-x')}\\
 \label{eq:G1_rewrite}    &
    +\sum_{i=1}^4\int_0^x\frac{\dx'}{x'-a_i}\frac{(1-2a_i)\Bigl((1-(1-a_i)a_i)m^4-(2(1-a_i)a_i+1)m^2S-(1-a_i)a_iS^2\Bigr)}{(m^2+S)^2\prod_{j=1,j\neq
            k}^4a_{ij}}     \\
      &=-i\pi\!+\!\frac{1}{2}\Bigl[
    2\Ef{-1}{\infty}{x}\!-\!\frac{1}{c_4}\Ef{0}{0}{x}\!-\!\Ef{-1}{0}{x}\!-\!\Ef{-1}{1}{x}\!-\!2\Ef{1}{0}{x}
\nonumber   \! -\!\Ef{1}{1}{x}\!+\!\sum_{i=1}^4\Ef{1}{a_i}{x}\Bigr]\,.
\end{align}
We apply this procedure to every MPL of weight one. Next, we differentiate MPLs of weight two. The derivative contains
MPLs of weight one, which we know how to express in terms of eMPLs. For example, we find,
\begin{equation}
    G(0,1;\chi) = -2\zeta_2+\int_0^x\dx'\frac{d\chi_{x'}}{dx'}\frac{G(1;\chi_{x'})}{\chi_{x'}}\,.
\end{equation}
Inserting~\eqn{eq:G1_rewrite} into the integrand, we find
\begin{align}
    G&(0,1;\chi) = \tfrac{1}{2c_4^2}\Ef{0,0}{0,0}{x}
    +\tfrac{1}{c_4}\Bigl[
    \tfrac{1}{2}\Ef{-1,0}{0,0}{x}\!+\!\tfrac{1}{2}\Ef{-1,0}{1,0}{x}\!-\!\Ef{-1,0}{\infty,0}{x}\!+\!\tfrac{1}{2}\Ef{0,-1}{0,0}{x}\!\\
    \nonumber&+\!\tfrac{1}{2}\Ef{0,-1}{0,1}{x}\!-\!\Ef{0,-1}{0,\infty}{x}
    +\!\tfrac{1}{2}\Ef{0,1}{0,1}{x}\!-\!\tfrac{1}{2}\Ef{0,1}{0,a_1}{x}\!-\!\tfrac{1}{2}\Ef{0,1}{0,a_2}{x}\!-\!\tfrac{1}{2}\Ef{0,1}{0,a_3}{x}\!\\
    \nonumber&-\!\tfrac{1}{2}\Ef{0,1}{0,a_4}{x}\!-\!\tfrac{1}{2}\Ef{1,0}{0,0}{x}\!+\!(\log x-i \pi)\Ef{0}{0}{x}
    \Bigr]
    +\tfrac{1}{2}\Ef{-1,-1}{0,0}{x}\!+\!\tfrac{1}{2}\Ef{-1,-1}{0,1}{x}\!\\
    \nonumber&-\!\Ef{-1,-1}{0,\infty}{x}\!+\!\tfrac{1}{2}\Ef{-1,1}{0,1}{x}\!-\!\tfrac{1}{2}\Ef{-1,1}{0,a_1}{x}\!-\!\tfrac{1}{2}\Ef{-1,1}{0,a_2}{x}\!-\!\tfrac{1}{2}\Ef{-1,1}{0,a_3}{x}
    -\!\tfrac{1}{2}\Ef{-1,1}{0,a_4}{x}\!\!\\
    \nonumber&+\!\tfrac{1}{2}\Ef{-1,-1}{1,0}{x}\!+\!\tfrac{1}{2}\Ef{-1,-1}{1,1}{x}\!-\!\Ef{-1,-1}{1,\infty}{x}\!+\!\tfrac{1}{2}\Ef{-1,1}{1,1}{x}-\!\tfrac{1}{2}\Ef{-1,1}{1,a_1}{x}\!-\!\tfrac{1}{2}\Ef{-1,1}{1,a_2}{x}
    \!\\
    \nonumber&-\!\tfrac{1}{2}\Ef{-1,1}{1,a_3}{x}\!-\!\tfrac{1}{2}\Ef{-1,1}{1,a_4}{x}\!-\!\Ef{-1,-1}{\infty,0}{x}-\!\Ef{-1,-1}{\infty,1}{x}\!+\!2\Ef{-1,-1}{\infty,\infty}{x}\!-\!\Ef{-1,1}{\infty,1}{x}\!\!\\
    \nonumber&+\!\Ef{-1,1}{\infty,a_1}{x}\!+\!\Ef{-1,1}{\infty,a_2}{x}
    +\!\Ef{-1,1}{\infty,a_3}{x}+\!\Ef{-1,1}{\infty,a_4}{x}\!-\!\tfrac{1}{2}\Ef{1,-1}{0,0}{x}\!-\!\tfrac{1}{2}\Ef{1,-1}{0,1}{x}\!\!\\
    \nonumber&+\!\Ef{1,-1}{0,\infty}{x}\!+\!\tfrac{1}{2}\Ef{1,1}{0,1}{x}\!-\!\tfrac{1}{2}\Ef{1,1}{0,a_1}{x}-\!\tfrac{1}{2}\Ef{1,1}{0,a_2}{x}
    -\!\tfrac{1}{2}\Ef{1,1}{0,a_3}{x}\!-\!\tfrac{1}{2}\Ef{1,1}{0,a_4}{x}\!\!\\
    \nonumber&-\!2(\log x-i
    \pi )\Ef{-1}{\infty}{x}\!-\!i \pi  \log x+\!\tfrac{\log ^2x}{2}\!+\!(\log x-i
    \pi)\Ef{-1}{0}{x}\!+\!(\log x-i \pi)\Ef{-1}{1}{x}\!\!-\!2\zeta_2\,.
\end{align}
The same idea can be applied to every MPL in \eqn{eqn:oepintegrand}. 
After this is done, we can easily perform the integral over $x$ 
using the definition of eMPLs in \eqn{eqn:ellPoly}. The final result reads
\begin{align}
\nonumber    S_{111}&(S,m^2)\Bigr|_{\dimeps^1}=\frac{1}{(m^2+S)c_4}\Bigl[
    \tfrac{1}{2}\log m^2\Bigl(4\Ef{1,0}{0,0}{1}\!-\!4\Ef{0,-1}{0,0}{1}\!-\!4\Ef{0,-1}{0,1}{1}+\!8\Ef{0,-1}{0,\infty}{1}
    \!\!\\
    \nonumber&-\!\tfrac{4}{c_4}\Ef{0,0}{0,0}{1}\Bigr)\!+\!\Ef{1,1,0}{0,0,0}{1}-\!2\Ef{0,-1,1}{0,0,1}{1}\!-\!2\Ef{0,-1,1}{0,1,1}{1}-\!2\Ef{0,-1,1}{0,\infty,1}{1}\!+\!5\Ef{0,1,-1}{0,0,0}{1}\!\!\\
    \nonumber&+\!5\Ef{0,1,-1}{0,0,1}{1}\!-\!4\Ef{0,1,-1}{0,0,\infty}{1}\!-\!5\Ef{0,1,1}{0,0,1}{1}+\!2\Ef{0,1,1}{0,0,a_1}{1}\!+\!2\Ef{0,1,1}{0,0,a_2}{1}\!+\!2\Ef{0,1,1}{0,0,a_3}{1}\!\!\\
    &+\!2\Ef{0,1,1}{0,0,a_4}{1}\!+\!3\Ef{0,1,-1}{0,1,0}{1}+\!3\Ef{0,1,-1}{0,1,1}{1}\!-\!6\Ef{0,1,-1}{0,1,\infty}{1}-\!2\Ef{0,1,-1}{0,a_1,0}{1}\!\!\\
    \nonumber&-\!2\Ef{0,1,-1}{0,a_1,1}{1}\!+\!4\Ef{0,1,-1}{0,a_1,\infty}{1}-\!2\Ef{0,1,-1}{0,a_2,0}{1}\!-\!2\Ef{0,1,-1}{0,a_2,1}{1}\!+\!4\Ef{0,1,-1}{0,a_2,\infty}{1}\!\!\\
    \nonumber&-\!2\Ef{0,1,-1}{0,a_3,0}{1}-\!2\Ef{0,1,-1}{0,a_3,1}{1}\!+\!4\Ef{0,1,-1}{0,a_3,\infty}{1}\!-\!2\Ef{0,1,-1}{0,a_4,0}{1}\!-\!2\Ef{0,1,-1}{0,a_4,1}{1}\!\!\\
    \nonumber&+\!4\Ef{0,1,-1}{0,a_4,\infty}{1}\!+\!2\Ef{1,0,-1}{0,0,0}{1}\!+\!2\Ef{1,0,-1}{0,0,1}{1}+\!2\Ef{1,0,-1}{0,0,\infty}{1}\!-\!3\Ef{1,0,1}{0,0,1}{1}\!+\!2\Ef{1,0,1}{0,0,a_1}{1}\!\!\\
    \nonumber&+\!2\Ef{1,0,1}{0,0,a_2}{1}\!+\!2\Ef{1,0,1}{0,0,a_3}{1}\!+\!2\Ef{1,0,1}{0,0,a_4}{1}\!+\!\zeta_2\Ef{0}{0}{1}\!+\!\tfrac{1}{c_4}\Bigl(\Ef{0,0,1}{0,0,1}{1}\!+\!2\Ef{0,1,0}{0,0,0}{1}\!\!\\
    \nonumber&+\!3\Ef{0,1,0}{0,1,0}{1}\!-\!2\Ef{0,1,0}{0,a_1,0}{1}\!-\!2\Ef{0,1,0}{0,a_2,0}{1}\!-\!2\Ef{0,1,0}{0,a_3,0}{1}\!-\!2\Ef{0,1,0}{0,a_4,0}{1}\!-\!\Ef{1,0,0}{0,0,0}{1}
    \Bigr)\Bigr]\,.
\end{align}
We see that, just like for the leading term in the expansion, we can cast the result in the form of a linear combination of eMPLs. It is easy to extend the procedure 
used to perform the integration at higher orders in $\dimeps$, which proves that the $S_{111}$ in $d=2-2\dimeps$ dimensions can be expressed in terms of E$_4$ functions at every order in dimensional regularization.

\section{The second master integral of the equal-mass sunrise}
\label{sec:secondmaster}

In this section we discuss the second master integral $S_{211}$ of the equal-mass sunrise topology.
We will follow the same strategy as for the first master integral, i.e., we start from the Feynman parameter representation
of $S_{211}$ and perform all the integrations sequentially in terms of eMPLs. In the process,
we will discover that the integration
kernels introduced in \eqn{eqn:firstkernels} are insufficient to perform all the integrals, and we will be forced to extend our set of integration kernels.

Let us start from the expression of the second master integral in terms of Feynman parameters:
\begin{equation}
  S_{211}(S,m^2)
  = \Gamma(2+2\dimeps)e^{2\gamma_E\dimeps}\int_0^{\infty}\dx_1\dx_2\dx_3\,
      \delta(1-x_3)\frac{x_1\mathcal{U}^{1+3\dimeps}}{\mathcal{F}^{2+2\dimeps}}\,,
\end{equation}
where the Symanzik-polynomials $\CF$ and $\CU$ have been defined in
\eqn{eqn:Symanzik}. The integral is finite in two dimensions, and the leading term in $\dimeps$ is
\begin{equation}
  \label{eqn:ellipticIntegral2}
    S_{211}(S,m^2)\Bigr|_{\dimeps^0}
=\int_0^{\infty}\dx_1\dx_2\dx_3\frac{\delta(1-x_3)x_1
    s_2}{[m^2(x_1+x_2+x_3)s_2+S\,s_3]^2}.
\end{equation}
The previous integral is very similar in structure to the Feynman parameter integral representation of
the first master integral in~\eqn{eq:SunriseEp0}. The main difference is that
we have an additional power of the polynomial~$\CF$ in the denominator and
pick up numerator terms from the polynomial~$\CU$. These will lead to the
appearance of rational prefactors in the integrand that need to be dealt
with using integration-by-parts identities.

To start, we proceed in exactly the same way as for $S_{111}$ in the previous section.
We eliminate the integration over $x_3$ using the $\delta$ function, and we
 map $x_1$ to the unit interval by changing variables according to to
$x_1=x/(1-x)$. Afterwards, we can factorize the quadratic polynomial in $x_2$
in the denominator and perform partial fractioning. The integral
over $x_2$ can then be done using standard techniques. We find the following representation of $S_{211}$:
\begin{equation}
S_{211}(S,m^2)\Bigr|_{\dimeps^0}=\int_0^1\dx\,(\CI_1+\CI_2)\,,
\label{eqn:sunrise2Xintegral}
\end{equation}
where
\begin{equation}\begin{split}
        \CI_1&=\frac{x(m^2(1-x+x^2)-S(1-x)x)}{m^2(m^2+S)^2(x-a_1)(x-a_2)(x-a_3)(x-a_4)}\,,\\
  \CI_2&=\frac{x(m^2(1-x+x^2)+S(1-x)x)}{(m^2+S)^3(x-a_1)(x-a_2)(x-a_3)(x-a_4)y}\log\chi\,,
  \label{eqn:theIs}
\end{split}\end{equation}
where $\chi$ and $y$ are defined in eqs.~\eqref{eqn:quantities} and~\eqref{eqn:y}. We will evaluate 
the integral in~\eqn{eqn:sunrise2Xintegral} by first computing a primitive of $\CI_i$, and then we evaluate the primitives
at the integration boundaries.

Let us begin by computing a primitive of $\mathcal{I}_1$. $\mathcal{I}_1$ does not contain any dependence
on the square root $y$, and so we can simply compute the primitive in terms of ordinary MPLs
after partial fractioning of the denominator.  Since we want to combine the
result with the elliptic contributions from $\mathcal{I}_2$, we find it convenient to write all ordinary MPLs
from the beginning in terms of eMPLs. We find,
\begin{equation}
\int\dx\,\CI_1 =
  \sum_{k=1}^4 
  \frac{a_k((1-(1-a_k)a_k)m^2-(1-a_k)a_kS)}{m^2 (m^2+S)^2\prod_{i=1,i\neq k}^4 a_{ki}}\Ef{1}{a_k}{x}\,.
\label{eqn:firstPrimitive}
\end{equation}

Next we turn to the second term in~\eqn{eqn:sunrise2Xintegral}. In contrast to
$\CI_1$, this term is of elliptic type due to the presence of $y$ in the denominator.  Let us recall that
$\log\chi$ can be cast in the form of a linear combination of eMPLs in~\eqn{eqn:logRewritten}. 
Partial fractioning in $x$ leads to integrals of the form
\begin{equation}
    \int\dx\frac{\CX(x)}{y\,(x-a_i)}\,,
    \label{eqn:ratInt}
\end{equation}
where $\CX(x)$ is some eMPL. At first glance, this integral looks very similar to \eqn{eqn:firstkernels} 
with $c=a_i$. 
There is however a crucial difference: the integral in~\eqn{eqn:firstkernels} has a simple pole at
$x=c$ for $c\neq a_i$. Since $x=a_i$ is a zero of the square root $y$, the integrand in~\eqn{eqn:ratInt} behaves like $(x-a_i)^{-3/2}$ for $x\sim a_i$. 
Hence, the point $x=a_i$ is not a pole in~\eqn{eqn:ratInt}, but rather a branch point of the square
root.
In the next subsection we discuss how to evaluate such integrals. 

\subsection{Intermezzo: Integration of eMPLs with rational coefficients}
\label{sec:intermezzo}
Let us begin by considering the simplest instance of the types of integrals encountered
in~\eqn{eqn:ratInt}, corresponding to $\CX(x)\equiv 1$.
Without loss of generality, we restrict the discussion to the case $i=1$.
We find that we can write the integral in~\eqn{eqn:ratInt} as,
\begin{equation}
    \label{eqn:basicIbp}
    \int\frac{\dx}{y (x-a_1)}
    = \frac{1}{a_{12}a_{13}a_{14}}\Bigl[-\frac{2y}{x-a_1}+(3a_1-\bar{s}_1)\int\frac{\dx}{y}(x-a_1)+2\int\frac{\dx}{y}(x-a_1)^2\Bigr],
\end{equation}
with $\bar{s}_n = s_n(a_2,a_3,a_4)$ and $s_n$ are the symmetric polynomials of degree $n$. 
We can immediately recognize our integration kernels
$\psi_0(0,x)$ and $\psi_{-1}(\infty,x)$ in the formula above. However, we also
find the integral
\begin{equation}
    \label{eqn:xsquareintegral}
    \int\frac{\dx}{y}x^2,
\end{equation}
which does not immediately translate into one of our kernels in~\eqn{eqn:firstkernels}, nor
can it be reduced to them using IBP identities. 
We are therefore forced to
introduce a new kernel, which we define by
\begin{equation}
    \label{eqn:Phi4tilde}
    \tilde{\Phi}_4(x)=\frac{1}{c_4 y}\Bigl(x^2-\frac{s_1}{2}x+\frac{s_2}{6}\Bigl).
\end{equation}
This kernel contains not only the necessary $x^2/y$ term, but also 
terms proportional to $\psi_{-1}(\infty,x)$ and $\psi_{0}(0,x)$. These terms are purely conventional at this point.

The object in \eqn{eqn:Phi4tilde} seems rather random at first sight; in
particular we do not seem to be consistent in our notation for the integration
kernels. There is a reason for not associating the name $\psi_i$ to
this kernel: we demand of our kernels $\psi_i$ that they have at most
simple poles.  The above kernel $\tilde{\Phi}_4(x)$, however, has a
double pole at $x=\infty$: indeed, letting $x=1/u$, we can
see that
\begin{equation}
    \dx \frac{x^2}{y} = -\frac{\du}{u^2}-\frac{s_1\du}{2u} + \CO(u^0),
\end{equation}
and consequently,
\begin{equation}
    \dx \,\tilde{\Phi}_4(x) = -\frac{\du}{u^2} + \CO(u^0)\,.
\end{equation}
Note that the term proportional to $x/y$ removes the simple pole at $u=0$,
so that $\tilde{\Phi}_4(x)$ has a double pole at infinity with vanishing residue.

As a consequence, $\tilde{\Phi}_4(x)$ does not meet the fundamental criterion that our integration kernels
should have at most simple poles.
However, a function with
a double pole gives rise to a primitive with a simple pole. We define a primitive of $\tilde{\Phi}_4$ by
\begin{equation}
    Z_4(x) = \int_{a_1}^x\dx'\,\Bigl(\tilde{\Phi}_4(x)+4c_4 \frac{\eta_1}{\omega_1}\frac{1}{y}\Bigr)\,,
    \label{eqn:Z4def}
\end{equation}
where $\omega_1$ is the first of
the two \emph{periods} of the elliptic curve, defined as 
\begin{equation}
    \label{eqn:ePeriod}
    \omega_1 = 2c_4 \int_{a_2}^{a_3}\frac{\dx}{y}=2\textrm{K}(\lambda)\qand\omega_2
    = 2c_4\int_{a_1}^{a_2}\frac{\dx}{y}=2i\textrm{K}(1-\lambda)\,,
\end{equation}
with
\begin{equation}
    \lambda = \frac{a_{14}a_{23}}{a_{13}a_{24}}.
\end{equation}
In the previous equation, $\textrm{K}(\lambda)\equiv\textrm{F}(1|\lambda)$ denotes the complete elliptic integral of the first kind.
Similarly, one defines the \emph{quasi-periods} of the elliptic curve by
\begin{align}
    \eta_1
    &= -\frac{1}{2}\int_{a_2}^{a_3}\dx\tilde{\Phi}_4(x)=\textrm{E}(\lambda)-\frac{2-\lambda}{3}\textrm{K}(\lambda)\,,\nnl
    \label{eqn:eQPeriod}
    \eta_2&=-\frac{1}{2}\int_{a_1}^{a_2}\dx\tilde{\Phi}_4(x)=-i\Bigl[\textrm{E}(1-\lambda)-\frac{1+\lambda}{3}\textrm{K}(1-\lambda)\Bigr]\,,
\end{align}
where $\textrm{E}(\lambda)\equiv \textrm{E}(1|\lambda)$ denotes the \emph{elliptic integral of the second kind}
\begin{equation}\label{eqn:elliptic_2nd_type}
    \textrm{E}(x|w^2) = \int_0^x\dt\frac{1-w^2t^2}{\sqrt{(1-t^2)(1-w^2t^2)}}.
    \end{equation}

Since $Z_4(x)$ is a primitive of a function with a double pole at infinity,
 $Z_4(x)$ itself only has a single pole at
$x=\infty$.  Consequently, we can use it to define new integration kernels
for our elliptic multiple polylogarithms,
\begin{equation}
    \label{eqn:z4kernel}
    \psi_{1}(\infty,x) = \frac{c_4}{y}Z_4(x)\,.
\end{equation}
Let us make an important comment about this kernel. Unlike the kernels considered in \eqn{eqn:firstkernels},
the function $\psi_{1}(\infty,x)$ is not rational or algebraic, but it is itself an integral that defines
a transcendental function. We see here one of the main differences between ordinary and elliptic polylogarithms: 
while ordinary MPLs only require integration kernels that are rational, this is no longer the case for elliptic polylogarithms.
Indeed, the kernel in~\eqn{eqn:z4kernel} cannot be
reduced to any of the kernels we defined so far. In particular, it is
independent of $\psi_{-1}(\infty,x)$, even though both kernels have a single
pole at infinity.

The fact that $\psi_{1}(\infty,x)$ is transcendental has far-reaching implications.
In particular, we will not be able to simplify powers of kernels or products with other kernels using partial fractioning. Instead, 
 powers or product of kernels that involve $Z_4$ will define new independent kernels.
 For example we can consider,
\begin{equation}
    \psi_{-2}(c,x) = \frac{y_c}{(x-c)y}Z_4(x)\,.
    \label{eqn:highKernels}
\end{equation}
Higher powers of $Z_4(x)$ will lead to further
kernels $\psi_{\pm n}$ with $n>2$. We obtain in this way an infinite tower of
independent integration kernels, which can be constructed in the above fashion by taking
combinations of powers of $Z_4(x)$ with elementary rational kernels.  This may
sound disconcerting at first sight: we set out to perform integrals over
functions involving elliptic square roots in a closed form only to find that
an infinite set of basis kernels is required. While there is no way around this
fact in general, it turns
out that for a given problem---concretely, a given number of
integrations---only a finite number of kernels will ever contribute. In fact,
for the problems discussed in this note the elementary kernels in~\eqn{eqn:firstkernels}, together with
the kernels defined in eqs.~\eqref{eqn:z4kernel} and~\eqref{eqn:high Kernels}, are
sufficient to perform all the integrals.

Finally, after this very long digression, we can return to the integral in \eqn{eqn:basicIbp}, and we can rewrite the term $x^2/y$ in
terms of $\tilde{\Phi}_4(x)$. We obtain
\begin{equation}\begin{split}\label{eq:psi1_reduce_result}
    \int&\frac{\dx}{y\,(x-a_1)} = \frac{2}{a_{12}a_{13}a_{14}}\Bigl[\frac{y}{x-a_1}
        +\Ef{0}{0}{x}\Bigl(\frac{3a_1^2-2a_1\,\bar{s}_1+\bar{s}_2}{6c_4}+4\frac{c_4\eta_1}{\omega_1}\Bigr)
        +c_4Z_4(x)
        \Bigr]\,.
\end{split}\end{equation}

\subsection{Back to the second master integral}
After this intermezzo on the integration on elliptic
curves, let us now return to the computation of the second master integral of the
sunrise topology with equal masses.  We had arrived at a point where we needed to perform
integrals of the type in~\eqn{eqn:ratInt}. For $\CX(x)\equiv 1$, we can simply follow the reasoning of the previous section.
In general, however, $\CX(x)$ will be an
elliptic polylogarithm. Fortunately, we can use IBP
to reduce all appearing integrals to our extended set of integral kernels. We
refer the reader to~\appref{app:ibp} for a more detailed discussion. Integrating the $\mathcal{I}_2$ term
in~\eqn{eqn:sunrise2Xintegral} comes down to finding primitives like the
following:
\begin{align}
    \int\frac{\dx}{y\,(x-a_1)}\Ef{0}{0}{x}&=\frac{1}{a_{12}a_{13}a_{14}}\Biggl[
    \frac{2y \Ef{0}{0}{x}}{a_1-x}
    +2 \int\frac{\dx}{x-a_1}y[\partial_x \Ef{0}{0}{x}]\\
    &+(3a_1-\bar{s}_1)\int\frac{\dx}{y}(x-a_1)\Ef{0}{0}{x}
    +2\int\frac{\dx}{y}(x-a_1)^2\Ef{0}{0}{x}\nonumber
    \Biggr].
\end{align}
The integrals appearing on the right-hand-side can be expressed in terms of our
kernels to yield
\begin{align}
    \int\frac{\dx}{x-a_1}y[\partial_x\Ef{0}{0}{x}]&=c_4\Ef{1}{a_1}{x}\,,\nnl
    \int\frac{\dx}{y}(x-a_1)\Ef{0}{0}{x}
    &= \Ef{-1,0}{\infty,0}{x}-\frac{a_1}{c_4}\Ef{0,0}{0,0}{x}\,,\nnl
    \int\frac{\dx}{y}(x-a_1)^2\Ef{0}{0}{x}
    &=c_4\Ef{1}{\infty}{x}+c_4Z_4(x)\Ef{0}{0}{x}-\frac{1}{2}(3a_1-\bar{s}_1)\Ef{-1,0}{\infty,0}{x}\nnl
    &+\Bigl[\frac{6a_1^2-a_1\,\bar{s}_1-\bar{s}_2}{6c_4}-\frac{4c_4\eta_1}{\omega_1}\Bigr]\Ef{0,0}{0,0}{x}\,.
\end{align}
In a similar fashion we can perform the remaining integrals contributing to the primitive of
$\mathcal{I}_2$ that are listed in~\appref{app:ints}. Combining them together with the primitive of $\mathcal{I}_1$
in~\eqn{eqn:firstPrimitive} we obtain the primitive of~\eqn{eqn:sunrise2Xintegral}:
\begin{align}
\CC_0\int\!\dx\,(\CI_1+\CI_2) &= 
-c_4\me{1}{x}-c_4\me{2}{x}+2c_4\me{3}{x}-\frac{2m^2}{m^2+S}\me{8}{x}\nnl
&\quad-\me{9}{x}+\mathcal{C}_1\Bigl[c_4\me{10}{x}+c_4\me{11}{x}-2c_4\me{12}{x}\nnl
&\quad+\me{13}{x}-c_4\me{14}{x}\Bigr]+c_4\bigl[\CC_2y+Z_4(x)\bigl]\me{4}{x}\nnl
&\quad+c_4\bigl[\CC_2y+Z_4(x)\bigl]\me{5}{x}-2c_4\bigl[\CC_2y+Z_4(x)\bigl]\me{6}{x}\nnl
&\quad+\bigl[\CC_2y+Z_4(x)+\CC_1c_4\log(x)\bigr]\me{7}{x}\,.
\label{eqn:primResult}
\end{align}
In the above expression we have used the coefficients $\mathcal{C}_i$
defined by
\begin{align}
    \mathcal{C}_0 &= m^2(S+9m^2)\,,\nnl
    \mathcal{C}_1 &= -4\frac{\eta_1}{\omega_1}-\frac{15m^4+12m^2S+S^2}{6(m^2+S)^2c_4^2}\,,\nnl
    \mathcal{C}_2
    &=-\frac{x^3(m^2+S)^2+x^2(3m^2-S)(m^2+S)-4m^2Sx-2m^4}{c_4\bigl[x^4(m^2+S)^2-2x^3(m^2+S)^2+x^2(3m^4+S^2)-2m^2x(m^2-S)+m^4\bigr]}\,.
\end{align}
Some comments about the structure of this result are in order. First of all,
note that the terms proportional to $\Ef{1}{a_i}{x}$ from the primitive of $\mathcal{I}_1$
in~\eqn{eqn:firstPrimitive} exactly cancel against corresponding terms in the
primitive of $\mathcal{I}_2$, so that the result has no logarithmic
singularities at roots of the elliptic curve.  
Second, let us note that the term proportional to $\log x\,\Ef{0}{0}{x}$ arises
from the shuffle identity
\begin{equation}
    \Ef{0,1}{0,0}{x}=\log x \,\Ef{0}{0}{x}-\Ef{1,0}{0,0}{x}\,.
\end{equation}
In order to obtain the final result for the second master of the sunrise, we
need to evaluate the primitive at the upper boundary of the integration region at $x=1$.
When we try to evaluate the primitive at $x=1$, we realize that some of the
eMPLs in the result have logarithmic singularities at $x=1$, namely the eMPLs
where the pole of the last integration is at $x=1$, these are
\begin{equation}
    \Ef{1}{1}{x}\,\quad\Ef{-1}{1}{x}\qand\Ef{-2}{1}{x}.
\end{equation}
The first of these eMPLs is already manifestly logarithmically divergent at $x=1$,
because $\Ef{1}{1}{x}=\log(1-x)$. The remaining two, however, do not have the divergent logarithms manifest. 
In order to be able to explicitly cancel the divergent logarithms and
to obtain a manifestly finite result, we need to extract the
divergence from the remaining two eMPLs.  The strategy for that is very simple:
we start from the integral
representation and subtract the pole at $x=1$. Let us illustrate this procedure on
$\Ef{-1}{1}{x}$. We have,
\begin{equation}
    \Ef{-1}{1}{x}  = \int_0^x\dx' \frac{y_1}{(x-1)y_{x'}}.
\end{equation}
The integrand has a pole at $x=1$ with unit residue. We subtract the pole, and then add it back,
\begin{align}
    \Ef{-1}{1}{x}
    &= \Ef{1}{1}{x}+\int_0^x\dx'\Bigl[\frac{y_1}{(x-1)y_{x'}}-\frac{1}{x-1}\bigr]\,.
\end{align}
The singularity is now manifest in the $\Ef{1}{1}{x}$ term, while the remaining
term is finite. We define
\begin{equation}
    \Eval{-1}{1} \equiv \int_0^1\dx'\Bigl[\frac{y_1}{(x-1)y_{x'}}-\frac{1}{x-1}\bigr],
\end{equation}
such that
\begin{equation}
    \label{eqn:regLim1}
    \lim_{x\to1} \bigl(\Ef{-1}{1}{x}-\Ef{1}{1}{x}\bigr) = \Eval{-1}{1}.
\end{equation}
Similarly we can regulate $\Ef{-2}{1}{x}$ using
\begin{equation}
    \Ef{-2}{1}{x}=\Ef{1}{1}{x}Z_4(1)+\int_0^x\dx'\Bigl[\frac{y_1Z_4(x)}{(x-1)y_{x'}}-\frac{Z_4(1)}{x-1}\Bigr]\,,
\end{equation}
and we obtain a finite limit
\begin{equation}
    \label{eqn:regLim2}
    \lim_{x\to1}\bigl(\Ef{-2}{1}{x}-\Ef{1}{1}{x}Z_4(1)\bigr) = \Eval{-2}{1}.
\end{equation}

Inserting the expressions for the regulated limits
from~\eqns{eqn:regLim1}{eqn:regLim2} into the primitive
in~\eqn{eqn:primResult}, we see that the logarithmic divergences cancel.
The final result for the second master integral reads 
\begin{align}
S_{211}(S,m^2)\Bigr|_{\dimeps^0}&=\frac{1}{\CC_0}\Bigl[
-c_4\me{1}{1}-c_4\Eval{-2}{1}+2c_4\me{3}{1}-\me{9}{1}\\
\nonumber&\,+\mathcal{C}_1\Bigl(c_4\me{10}{1}+c_4\me{11}{1}-2c_4\me{12}{1}+\me{13}{1}-c_4\me{14}{1}\Bigr)\\
\nonumber&\,+\Bigl(Z_4(1)-2\frac{y_1}{c_4}\Bigl)\Bigl(c_4\me{4}{1}+c_4\Eval{-1}{1}-2c_4\me{6}{1}+\me{7}{1}\Bigr)
\Bigr]\,.
    \end{align}
While we only present the explicit result for the leading order in the $\dimeps$ expansion, 
it is easy to repeat the steps outlined in this section 
to obtain analytic results at every order in $\dimeps$ 
in terms of eMPLs and $Z_4$.

\section{A lightning summary of elliptic polylogarithms}
\label{sec:recap}

In the previous sections we have shown that all the members of the sunrise
topology can be expressed in terms of the eMPLs defined in
eqs.~\eqref{eqn:ellPoly}, \eqref{eqn:firstkernels} and~\eqref{eqn:psi1}. We
discuss how to extend these results to members of the sunrise topology
depending on three different masses in the next section.  The goal of this section is to present a concise summary  of the main concepts
introduced in previous sections, and to shortly discuss some further properties
of eMPLs.  For a more detailed discussion of these functions and their
properties, we refer to ref.~\cite{longelliptic}.

\phantom{a}

\paragraph{\underline{\it{Elliptic curves and their invariants:}}} Our main objects of interest are
    integrals whose integrands depend on a
    rational function of the integration variable $x$ as well as of the square
    root $y$ of a quartic polynomial in $x$,
    \begin{equation} y=\sqrt{(x-a_1)(x-a_2)(x-a_3)(x-a_4)}\,. \end{equation}
    The variables $x$ and $y$ define an elliptic curve, and so we actually look
    at integrals on an elliptic curve. The quantity $\vec{a}=(a_1,a_2,a_3,a_4)$
    is called the \emph{root vector} of the elliptic curve. There are certain
    `invariants' attached to an elliptic curve, the \emph{periods} $\omega_i$
    and the \emph{quasi-periods} $\eta_i$, cf. eqs.~\eqref{eqn:ePeriod}
    and~\eqref{eqn:ePeriod}. The periods and quasi-periods are functions only
    of the root vector, and can be expressed in terms of the elliptic integrals
    of the first and second kind K and E, cf. eqs.~\eqref{eqn:ellK}
    and~\eqref{eqn:elliptic_2nd_type}. Note that a cubic polynomial also
    defines an elliptic curve. We do not discuss the case of a cubic polynomial
    in this paper, but we refer to~\cite{longelliptic}.
    
    \phantom{a}
    
\paragraph{\underline{\it{IBP identities:}}} An integral on an elliptic curve can be
    reduced via IBP identities to a set of basis/master integrals. The relevant IBP identities are collected in Appendix~\ref{app:ibp}. The
    kernels corresponding to the master integrals read 
    \begin{equation}\label{eq:final_quartic}
     \begin{split}
      \psi_0(0,x) &\,= \frac{c_4}{y}\,,\\
      \psi_1(c,x)&\, = \frac{1}{x-c}\,,\qquad \psi_{-1}(c,x) = \frac{y_c}{y(x-c)}\,, \\
      \psi_1(\infty,x)&\, = \frac{c_4}{y}\,Z_4(x)\,,\qquad \psi_{-1}(\infty,x) = \frac{x}{y}\,, \\
      \psi_{-n}(\infty,x)&\, = \frac{x}{y}\,Z_4^{(n-1)}(x)-\frac{\delta_{n2}}{c_4}\,,\\
       \psi_n(c,x)&\, = \frac{1}{x-c}\,Z_4^{(n-1)}(x)-\delta_{n2}\,\big(\tilde{\Phi}_4(x)+4c_4\frac{\eta_1}{\omega_1}\frac{1}{y}\big)\,,\\
      \psi_n(\infty,x) &\,= \frac{c_4}{y}\,Z_4^{(n)}(x)\,,\qquad \psi_{-n}(c,x)  = \frac{y_c}{y(x-c)}\,Z_4^{(n-1)}(x)\,.
     \end{split}
    \end{equation} 
    The quantities $Z_4$ and $\tilde{\Phi}_4$ are defined in eqns.~(\ref{eqn:Z4def}) and~(\ref{eqn:Phi4tilde}).
    The integration kernels are chosen such that they have at most simple poles, and so their integrated versions
     have at most logarithmic singularities. The quantities $Z_4^{(n)}$ are polynomials in $Z_4$ of the form
    \begin{equation}
    Z_4^{(n)}(x) = \frac{(-1)^n}{2^nn!}\,Z_4(x)^n + \ldots\,,
    \end{equation}
    with $Z^{(1)}_4(x)=Z_4(x)$, and the dots indicate terms that involve fewer powers of $Z_4$, such that they cancel the pole of order $n$ at $x=\infty$
    of $Z_4(x)^n$. In the context of the sunrise integrals only the case $n=1$ is relevant. We do therefore not discuss the case $n>1$ here, but we refer to the literature
    for a detailed discussion~\cite{longelliptic}.

\phantom{a}
    
\paragraph{\underline{\it{Elliptic polylogarithms:}}} Integrating multiple times over the elliptic integration kernels in~\eqn{eq:final_quartic},
    leads to \textit{multiple elliptic polylogarithms} (eMPLs)
    \begin{equation}
        \label{eqn:ellPoly2}
      \Ef{n_1,\ldots,n_k}{c_1,\ldots,c_k}{x}=\int_0^x\dt\,\,\psi_{n_1}(c_1,t) \Ef{n_2,\ldots,n_k}{c_2,\ldots,c_k}{t}\,,
    \end{equation}
    where $n_i\in\ZZ$, $c_i\in \ZC\cup\lbrace\infty\rbrace$ and the recursion
    starts with $\EE(;x)=1$. $k$ is called the \emph{length} and $n_1+\ldots+n_k$ is the \emph{weight}. Since the kernels have at most simple poles, eMPLs have at most logarithmic singularities. 
 
 \phantom{a}
 
\paragraph{\underline{\it{Shuffle algebra:}}} eMPLs form a shuffle algebra
    graded by the length. In other words, a product of two eMPLs can be recast in
    the form of a linear combination of eMPLs, \begin{equation} \textrm{E}_4(\vec
      d_1;x)\,\textrm{E}_4(\vec d_2;x) \sum_{\vec d\in \vec d_1\shuffle\vec
      d_2}\textrm{E}_4(\vec d;x)\,, \end{equation} with $\vec d_1 =
    \left(\begin{smallmatrix}n_1&\ldots& n_k\\ c_1&\ldots&
    c_k\end{smallmatrix}\right)$ and similarly for $\vec d_2$, and the sum runs
    over all shuffles of $\vec d_1$ and $\vec d_2$, i.e., all possible permutations
    of $\vec d_1\cup\vec d_2$ that preserve the relative orderings within $\vec
    d_1$ and $\vec d_2$.

\phantom{a}

\paragraph{\underline{\it{Completeness and independence of the integration
    kernels:}}} The integration kernels in~\eqn{eq:final_quartic} define a complete
    and independent set. In particular, eMPLs are linearly independent (for generic
    values of $x$).  The set of kernels is infinite, which is a feature of the
    elliptic curve. In general, for every $c\in \mathbb{C}\cup\{\infty\}$ with
    $c\neq a_i$, there are two infinite towers of integration kernels $\psi_{\pm
    n}(c,x)$, $n\ge 1$. If $c=a_i$, then we can always reduce the tower with
    negative index via IBP identities (cf. \secref{sec:intermezzo} for an
    illustration of  this reduction in the case of $\psi_{-1}(a_i,x)$, in
    particular~\eqn{eq:psi1_reduce_result}). For the solution of a particular
    integral, however, only a finite number of kernels is required.  Finally, every
    integral involving eMPLs and rational functions involving $x$ and $y$ can
    be performed in terms of the same class of functions. For an explicit algorithm
    how to perform such integrals, we refer to Appendix~\ref{app:ibp} and
    ref.~\cite{longelliptic}.
    
\phantom{a}

\paragraph{\underline{\it{Relationship to ordinary MPLs and elliptic
    integrals:}}} eMPLs contain various classes of special functions that are
    well-known in the mathematics and physics literature. In
    particular,~\eqn{eqn:psi1} implies that ordinary MPLs are a special case of
    eMPLs, \begin{equation}\label{eq:E_to_G_4} \Ef{1&\ldots&1}{c_1& \ldots& c_k}{x}
    = G(c_1,\ldots,c_k;x)\,,\qquad c_i\neq \infty\,.  \end{equation} In addition,
    the incomplete elliptic integrals of the first, second and third kinds are
    special cases of eMPLs. For example, if the root vector is $\vec a =
    (-1/w,-1,1,1/w)$, $0<w<1$, we have
    \begin{equation}\begin{split}\label{eq:FEPi}
            \textrm{F}(x|w^2) &\,= \int_0^x\frac{dt}{\sqrt{(1-t^2)(1-w^2t^2)}}
            =\frac{2}{1+w}\Ef{0}{0}{x}\,,\\
            \textrm{E}(x|w^2) &\,= \int_0^xdt\,\frac{1-w^2t^2}{\sqrt{(1-t^2)(1-w^2t^2)}} \\
            &\,= 
            \Biggl[2(1+w)\frac{\eta_1}{\omega_1}+\frac{5-w^2}{3(1+w)}\Biggr]\Ef{0}{0}{x}
            -\frac{1}{2}(1+w)\Bigl(Z_4(x)-Z_4(0)\Bigr)\,,\\
            \Pi(n^2,x|w^2) &\,= \int_0^x\frac{dt}{\sqrt{(1-t^2)(1-w^2t^2)}}\,\frac{1}{1-n^2t^2} =
            \frac{1}{2nw}\Biggl(\frac{\Ef{-1}{-1/n}{x}}{y_{-1/n}}-\frac{\Ef{-1}{1/n}{x}}{y_{1/n}}\Biggr)\,.
    \end{split}\end{equation}

\phantom{a}
    
\paragraph{\underline{\it{Relationship to the eMPLs by Brown \& Levin:}}}
    Finally, our eMPLs are closely related to multiple elliptic polylogarithms that
    appear in pure mathematics and string
    theory~\cite{BrownLevin,Broedel:2014vla,MatthesThesis}. In particular, in
    ref.~\cite{longelliptic} we show that every E$_4$ function can be written as a
    linear combination of the elliptic polylogarithms of
    ref.~\cite{BrownLevin,Broedel:2014vla,MatthesThesis}, and
    vice-versa (Up to a technical distinction which is irrelevant in the
    context of this paper; see ref.~\cite{longelliptic} for details). In other
    words, the functions defined in \eqn{eqn:ellPoly} are simply an alternative
    basis for the eMPLs in ref.~\cite{BrownLevin,Broedel:2014vla,MatthesThesis}. As
    such, our functions deserve indeed to be called \emph{elliptic polylogarithms}.

\section{Several applications}
\label{sec:application}

In this section we present additional applications of our elliptic
polylogarithms to integrals from the sunrise topology. While in previous
sections we have focused on the equal-mass case, from now on we consider the
generic case with three different non-zero masses.  Using IBP identities, one
finds four master integrals with three propagators for the sunrise topology,
which we may choose as
\begin{equation}
S_{111}\,,\qquad S_{211}\,,\qquad S_{121}\,\qand S_{112}\,.
\end{equation}
The three master integrals with an additional power on one of the propagators
are obviously related by a simple permutation of the masses $m_i$. It will
therefore be sufficient to discuss the integrals $S_{111}$ and $S_{211}$, and
the remaining integrals can be obtained by symmetry.  For this reason, we will
again refer to $S_{111}$ and $S_{211}$ as the \emph{first} and \emph{second}
master integrals of the sunrise topology. 

Analytic results for the first master integral $S_{111}$ with three different
masses can be found in ref.~\cite{Adams:2014vja,Adams:2015gva}, while one of the main results of this
paper is to present an analytic expression for $S_{211}$ with different masses
for the first time.  In order to calculate this result, we take an approach
that is complementary to the Feynman parameters used in the equal-mass case: we
compute analytic expressions for $S_{111}$ and $S_{211}$ from dispersion
relations.  In this way we demonstrate the flexibility of our language of
elliptic polylogarithms, and at the same time we obtain more compact
representations for the sunrise integrals in the case of three different
masses~\footnote{We have also performed the computations in the different-mass
  case from Feynman parameters. While both approaches give the same results
  numerically, the results obtained for $S_{211}$ with different masses leads
  to a rather complicated and lengthy analytic expression.}.

To compute the sunrise integrals in the dispersive approach, we begin by first
computing the imaginary part of the master integrals. The imaginary part can be
computed using the optical theorem by replacing all propagators of the sunrise
diagram by on-shell $\delta$ functions, and we discuss in detail the
computation of the maximal cut of $S_{111}$ in terms of eMPLs. Once the
imaginary part is obtained, we can recover the master integrals $S_{111}$ and
$S_{211}$ by performing a dispersion integral.

\subsection{The maximal cut of the first master integral}
\label{sec:sunrisecut}

In this section we discuss the computation of the maximal cut of $S_{111}$.
In \rcite{Remiddi:2016gno} it was shown that the imaginary part of $S_{111}$ can be written as follows
\begin{equation}
\label{eqn:ImSunrise}
 \Im S_{111}(s,m_1^2,m_2^2,m_3^2) = 
 S_{\dimeps} \int_{(m_2+m_3)^2}^{(\sqrt{s}-m_1)^2}\dx\frac{\Bigl(\frac{R_2(x,m_2^2,m_3^2)R_2(s,x,m_1^2)}{sx}\Bigr)^{(d-2)/2}}{\sqrt{R_2(x,m_2^2,m_3^2)R_2(s,x,m_1^2)}} \,,
\end{equation}
with $s=-S=p^2$ and where we defined
\begin{equation}
    S_{\dimeps} = \frac{2^d\pi^{3/2}}{\Gamma(3-\frac{d}{2})\Gamma(\frac{d-1}{2})} \qquad \mbox{with}
    \qquad \lim_{d \to 2} S_\dimeps \to 4 \pi \,, \label{eqn:Seps}
\end{equation}
and the K\"allen function is
\begin{equation}
R_2(x,y,z) = x^2+y^2+z^2-2xy-2xz-2yz.
\end{equation}
Since the K\"allen function is a polynomial of degree two, we can immediately recognize the square root of a
quartic polynomial in the denominator of~\eqn{eqn:ImSunrise}. The quartic polynomial is
given by
\begin{equation}
  y^2\equiv R_2(x,m_2^2,m_3^2)R_2(s,x,m_1^2)=(x-b_1)(x-b_2)(x-b_3)(x-b_4),
\end{equation}
where the roots are given by
\begin{equation}
b_1 = (m_2-m_3)^2,\quad b_2 = (m_2+m_3)^2,\quad b_3 = (m_1-\sqrt{s})^2\quad\textrm{and}\quad b_4 = (m_1+\sqrt{s})^2. \label{eqn:rootscut}
\end{equation}
For $s > (m_1+m_2+m_3)^2$, the roots are real and ordered as $b_1 <
b_2 < b_3 < b_4$, which we assume in the following.  We also assume that the
three masses $m_i$ are distinct and non-vanishing, so that no two roots
coincide.  

We can now rewrite \eqn{eqn:ImSunrise} manifestly as integral on an elliptic
curve,
\begin{align}\label{eq:Im_S111_int}
    \Im S_{111}(s,m_1^2,m_2^2,m_3^2) &\,= S_{\dimeps} \int_{b_2}^{b_3}\frac{\dx}{y}\Bigl(\frac{R_2(x,m_2^2,m_3^2)R_2(s,x,m_1^2)}{s x}\Bigr)^{(d-2)/2}\\
    \nonumber &\,=S_{\dimeps}\,\bigl(C_{111}^{(0)}(s,m_1^2,m_2^2,m_3^2) + \dimeps\,C_{111}^{(1)}(s,m_1^2,m_2^2,m_3^2)+\mathcal{O}(\dimeps^2)\bigl)\,.
\end{align}
The leading term in $\dimeps$ is given by
\begin{align}
C_{111}^{(0)}(s,m_1^2,m_2^2,m_3^2) = \int_{b_2}^{b_3}\frac{\dx}{y} = \int_{b_2}^{b_3}\dx\,\frac{\psi_0(0,x)}{c_4}=
    \frac{1}{c_4}\Bigl[\Ef{0}{0}{b_3}-\Ef{0}{0}{b_2}\Bigr]\,,
    \end{align}
    where in the second step we have recognized the integration kernel $\psi_0(0,x)$.
The coefficient $C_{111}^{(1)}$ can be represented by the integral
\begin{align}
\nonumber C_{111}^{(1)}&(s,m_1^2,m_2^2,m_3^2) = \int_{b_2}^{b_3}\!\!\frac{\dx}{y} \left[\log\left({R_2(x,m_2^2,m_3^2)R_2(s,x,m_1^2)}\right)-\log s-\log x\right]\\
&\, = \int_{b_2}^{b_3}\!\!\frac{\dx}{y} \left[-\log s-\log x+\log(b_1b_2b_3b_4) + \sum_{i=1}^4\log\left(1-\frac{x}{b_i}\right)\right]\\
\nonumber&\,=\frac{1}{c_4}\!\int_{b_2}^{b_3}\!\!\!\dx\,\psi_{0}(0,x)\Bigl[
\Ef{1}{b_1}{x}\!+\!\Ef{1}{b_2}{x}\!+\!\Ef{1}{b_3}{x}\!+\!\Ef{1}{b_4}{x}\!-\!\Ef{1}{0}{x}\!+\!\frac{1}{c_4}\log\Bigl(\frac{b_1b_2b_3b_4}{s}\Bigr)
\Bigr]\,,
\end{align}
where in the last step we have used the fact that ordinary MPLs are a subset of
\mbox{eMPLs}, cf.~\eqn{eq:E_to_G_4}.
The remaining integral can now be performed by using the recursive definition of eMPLs in \eqn{eqn:ellPoly}, with the result
\begin{align}
      \nonumber  C_{111}^{(1)}&(s,m_1^2,m_2^2,m_3^2)
        =\frac{1}{c_4}\Biggl(\Ef{0,1}{0,b_1}{b_3}-\Ef{0,1}{0,b_1}{b_2}
        + \Ef{0,1}{0,b_2}{b_3}-\Ef{0,1}{0,b_2}{b_2}+ \Ef{0,1}{0,b_3}{b_3}\\
        &\,-\Ef{0,1}{0,b_3}{b_2}
        + \Ef{0,1}{0,b_4}{b_3}-\Ef{0,1}{0,b_4}{b_2}\Biggr)+ \log\Bigl(\frac{b_1b_2b_3b_4}{s}\Bigr)\Bigl(\Ef{0}{0}{b_3}-\Ef{0}{0}{b_2}\Bigr)\,.
\end{align}
Let us conclude this section by commenting on how the results of this section
generalize to higher orders in $\dimeps$. From~\eqn{eq:Im_S111_int} we see that
the integrand at any order contains products of logarithms only. Using the
shuffle algebra properties of MPLs and eMPLs, and the fact that ordinary MPLs are
a subset of eMPLs, we can see that the integrand can always be written in terms
of E$_4$ functions. The resulting integrals can then always easily be performed
in terms of eMPLs, proving that the maximal cut of the sunrise integral can be
expressed in terms of eMPLs to all orders in dimensional regularization.

\subsection{The first master integral from a dispersion relation}
\label{sec:dispersion}

In this section we present the computation of the master integrals $S_{111}$ and $S_{211}$ using a dispersive approach.  As we will
see, this allows us to obtain a very compact representation for all master
integrals, even in the unequal mass case.

Let us go back to the formula for the imaginary part of the two-loop massive
sunrise with different masses and for general values of the dimensions $d$,
\eqn{eqn:ImSunrise}.  We can reconstruct the full sunrise integral through a
dispersion relation as follows 
\begin{align}
S_{111}&(s,m_1^2,m_2^2,m_3^2) = 
\frac{1}{\pi}\, \int_{m_{123}^2}^\infty \frac{du}{u-s}\, \Im S_{111}(u,m_1^2,m_2^2,m_3^2)  \\
\nonumber&\,= \frac{S_\epsilon}{\pi}  \int_{m_{123}^2}^\infty \frac{du}{u-s} 
\int_{(m_2+m_3)^2}^{(\sqrt{s}-m_1)^2} \frac{dt}{\sqrt{R_2(t,m_2^2,m_3^2) R_2(u,t,m_1^2) }}
\left( \frac{R_2(t,m_2^2,m_3^2) R_2(u,t,m_1^2)}{u\, t}\right)^{(d-2)/2}
\end{align}
where $S_\dimeps$ was defined in \eqn{eqn:Seps}, $m_{123}^2 = (m_1+m_2+m_3)^2$ and we omitted the small imaginary part associated to $s$, $s \to s+ i 0^+$.
As a first step, we exchange the integrals and write
\begin{align}
S_{111}(s,m_1^2,m_2^2,m_3^2) &= \frac{S_{\epsilon}}{\pi} \int_{(m_2+m_3)^2}^{\infty} \frac{dt}{\sqrt{R_2(t,m_2^2,m_3^2)}} 
\left( \frac{R_2(t,m_2^2,m_3^2)}{t} \right)^{(d-2)/2} \nonumber \\
&\qquad \times \int_{(\sqrt{t}+m_1)^2}^\infty \frac{du}{u-s} \frac{1}{\sqrt{R_2(u,t,m_1^2)}}
\left( \frac{R_2(u,t,m_1^2)}{u} \right)^{(d-2)/2} \label{eqn:sundisp1}\,.
 \end{align}
Upon inspecting \eqn{eqn:sundisp1}, since the integral over
$u$ contains only one quadratic square root,  it is obvious that this integral can easily be performed in
terms of multiple polylogarithms at every order in $\dimeps$. The result can then be integrated over $t$ in terms of our $\EE$ functions.  For simplicity, we limit ourselves to
the leading order in $\epsilon$, though we stress that there is no conceptual obstacle to extend the result to higher orders.  Furthermore, we assume without loss of generality that the
masses are ordered according to $m_3<m_2<m_1$, and we work below
threshold to deal with real quantities only, i.e., $0<s < m_{123}^2$.
We find
\begin{equation}\begin{split}
S_{111}&(s,m_1^2,m_2^2,m_3^2) = 4\, \int_{(m_2+m_3)^2}^{\infty} \frac{dt}{\sqrt{R_2(t,m_2^2,m_3^2)}}   \int_{(\sqrt{t}+m_1)^2}^\infty \frac{du}{u-s} \frac{1}{\sqrt{R_2(u,t,m_1^2)}}
 \label{eqn:sundisp2}\,.
 \end{split}\end{equation}
We can now perform explicitly the integral over $u$ and we find
\begin{align}
\!\!\!\!S_{111}&(s,m_1^2,m_2^2,m_3^2) = 2\! \int_{(m_2+m_3)^2}^{\infty}\!\! \frac{dt}{\sqrt{R_2(t,m_2^2,m_3^2) R_2(s,t,m_1^2)}} 
\ln\left( \frac{t+m_1^2-s + \sqrt{R_2(s,t,m_1^2)}}{t+m_1^2-s - \sqrt{R_2(s,t,m_1^2)}}\right).
 \end{align}
Finally, we change of variables according to 
\begin{equation}
t = b_3 + \frac{(b_4-b_3)}{4}
\frac{(1+x)^2}{x}\,,
\end{equation}
where the four roots $b_1$, $b_2$, $b_3$, $b_4$ are
defined in \eqn{eqn:rootscut}, and we are left with
\begin{equation}\begin{split}
\!\!\!\!S_{111}&(s,m_1^2,m_2^2,m_3^2) =\frac{2}{ \sqrt{s}\, m_1}\int_{0}^{Q_2} \frac{dx}{\sqrt{(x-Q_1)(x-1/Q_1)(x-Q_2)(x-1/Q_2)}} 
\ln\left( \frac{q + x }{x(\, 1+ x\, q)}\right) \,,\label{eqn:sundisp3}
 \end{split}\end{equation}
where we defined
\begin{equation}
  q = \frac{\sqrt{s}}{m_1}\,, \qquad Q_1 =  \frac{\sqrt{a_1-a_3} - \sqrt{a_1-a_4}}{\sqrt{a_1-a_3}+\sqrt{a_1-a_4}} \,, \qquad
 Q_2 =  \frac{\sqrt{a_2-a_3} - \sqrt{a_2-a_4}}{\sqrt{a_2-a_3}+\sqrt{a_2-a_4}} \,.
\end{equation}
It is now entirely straightforward  to integrate \eqn{eqn:sundisp3} in terms of
our $\EE$ functions. We recast the logarithms in the form of eMPLs
functions, and we obtain
\begin{align}
\ln\left( \frac{q + x }{x(\, 1+ x\, q)}\right) &= \ln{q} + \ln{(1+ x/q)} - \ln{x} - \ln{(1 + x\,q)} \nonumber \\
&= \ln{q} + \Ef{1}{-q}{x} - \Ef{1}{0}{x} - \Ef{1}{-1/q}{x}\,,
\end{align}
which immediately yields
\begin{align}
S_{111}(s,m_1^2,m_2^2,m_3^2) = \frac{4 \sqrt{2}}{\sqrt{P_0}}\, F_1^{(1)}(s,m_1^2,m_2^2,m_3^2)
\end{align}
with the definitions
\begin{align}
P_0 &=  2 m_1^2 m_2^2  + 2m_1^2 m_3^2 + 2 m_2^2 m_3^2  
+ 2 \,(m_1^2  + m_2^2  + m_3^2 ) \, s -(m_1^4 + m_2^4 + m_3^4) - s^2  \nonumber \\
&- \sqrt{a_1-a_3}\sqrt{a_1-a_4}\sqrt{a_2-a_3}\sqrt{a_2-a_4}\,,
\end{align}
and
 \begin{equation}\begin{split}
F_1^{(1)}&(s,m_1^2,m_2^2,m_3^2)=\Ef{0,1}{0,0}{Q_2} + \Ef{0,1}{0,-1/q}{Q_2}  - \Ef{0,1}{0,-q}{Q_2} - \ln{q} \Ef{0}{0}{Q_2}  \,. \label{eqn:sundisp4}
 \end{split}\end{equation}
  
\subsection{The second master integral from a dispersion relation}
In this section we extend the results of the previous section to the second master integral of the 
sunrise integrals with different masses. Without loss of generality, we discuss $S_{121}$, and the other cases are obtained by
symmetry. In addition, the second master integral is related to the first master integral by differentiation,
\begin{align}
S_{121}(s,m_1^2,m_2^2,m_3^2) = \frac{\partial}{\partial m_2^2} S_{111}(s,m_1^2,m_2^2,m_3^2)\,.
\end{align}

Hence, we can obtain an integral representation of $S_{121}$ by simply differentiating with respect to the mass
in the dispersive representation of $S_{111}$ in 
\eqn{eqn:sundisp3}.  
However, upon differentiating
under the integral sign, we generate a potential end-point singularity in the upper
integration limit $x \to Q_2$, which we need to regularize. We write
\begin{align}
 \, S_{121}(s,\,&m_1^2,m_2^2,m_3^2)   =  \frac{\partial}{\partial m_2^2} \, 
 S_{111}(s,m_1^2,m_2^2,m_3^2)  
 \\
&= \frac{2}{\sqrt{s}\, m_1}  \int_{0}^{Q_2-\delta} \frac{dx}{\sqrt{(x-Q_1)(x-1/Q_1)(x-Q_2)(x-1/Q_2)}} \,
\ln\left( \frac{q + x }{x(\, 1+ x\, q)}\right) \nonumber \\
 & \qquad \qquad \quad \times
\frac{x\, \left( m_1 \sqrt{s} + (m_1^2 -m_2^2 +m_3^2 +s )\, x + m_1 \sqrt{s} \, x^2 \right) }{m_1^2 \, s
(x-Q_1)(x-1/Q_1)(x-Q_2)(x-1/Q_2)} \nonumber \\
\nonumber&+ \frac{2}{\sqrt{s}\, m_1}  \frac{1}{\sqrt{-\delta}}\frac{1}{\sqrt{(Q_2-Q_1)(Q_2-1/Q_1)(Q_2-1/Q_2)}} \,
\ln\left( \frac{q + Q_2 }{Q_2(\, 1+ Q_2\, q)}\right) \,,
\end{align}
where we introduced a small regularization parameter $\delta$.  If we perform partial
fractioning in $x$ and use the expression of the logarithms in terms of
$\EE$ functions, we immediately see that we are left with the following types
of integrals to do
\begin{align}
 \int_{0}^{Q_2-\delta} \frac{dx}{y}  \frac{1}{x-Q}  \Ef{n}{c}{x}\,,\qquad Q\in\{1/Q_1,Q_1,Q_2,1/Q_2\}\,,
\end{align}
for different values of $(n,c)$. The integrals above can be
performed in a straightforward way in terms of $\EE$ functions using the techniques introduced in previous sections.
In the final result, all poles in $\delta$ cancel, and we can safely
take the limit $\delta \to 0$. We
find that $S_{121}(s,m_1^2,m_2^2,m_3^2)$ can be written in a very compact form as
follows
\begin{align}
 S_{121}(s,m_1^2,m_2^2,m_3^2) &= 
\frac{1}{D\,\sqrt{2\, P_0\,}}  \left[ F_{2}^{(-1)}(s,m_1^2,m_2^2,m_3^2) 
- \frac{4}{3} P_3\, F_1^{(1)}(s,m_1^2,m_2^2,m_3^2) \right] \nonumber \\
&+ \frac{\sqrt{2\, P_0}}{D} \left[ F_{2}^{(1)}(s,m_1^2,m_2^2,m_3^2) 
- 2 \frac{\eta_1}{\, \omega_1} P_1\, S(s,m_1^2,m_2^2,m_3^2) \right] \nonumber \\
&+ \frac{1}{D}F_2^{(0)}(s,m_1^2,m_2^2,m_3^2)
\end{align}
where we introduced the further abbreviations
\begin{align}
&\mu_1 = -m_1+m_2+m_3\,,\quad \mu_2 = m_1-m_2+m_3\,,\quad \mu_3 = m_1+m_2-m_3\,,
\nonumber \\
&D = m_2^2 (s-m_{123}^2)(s-\mu_1^2)(s-\mu_2^2)(s-\mu_3^2)\,, 
\end{align}
and
 \begin{align}
 F_2^{(-1)}(s,m_1^2,m_2^2,m_3^2) = - \frac{4}{3}\, P_2 \, \Ef{0}{0}{Q_2}
 \end{align}
 \begin{align}
 F_2^{(0)} (s,m_1^2,m_2^2,m_3^2) = &- P_4 \, \Ef{-1}{\infty}{Q_2} + P_5\, \Ef{-1}{-1/q}{Q_2} 
 - P_6 \, \Ef{-1}{-q}{Q_2}  \nonumber \\&+ P_7\, \Ef{-1}{0}{Q_2} - \ln{\frac{q}{Q_2}}
 \end{align} 
 \begin{align}
 F_2^{(1)}(s,m_1^2,m_2^2,m_3^2)  = P_1 &\left[ -\frac{1}{2}\, \Ef{2}{0}{Q_2} - \frac{1}{2} \Ef{2}{-1/q}{Q_2} + \frac{1}{2} \Ef{2}{-q}{Q_2} \right.  \\
\nonumber &\left.\; 
 + \frac{2 \eta_1}{\omega_1}\, \Ef{0}{0}{Q_2} - \frac{i \pi}{\omega_1}\, \ln{ \left( \frac{(q+Q_2)}{Q_2(1 + q \, Q_2)} \right) } 
 + \frac{1}{2} \ln{\left( \frac{q}{Q_2} \right)} 
 Z_4^{(1)}(0)  \right]
\end{align}
The polynomials $P_i$ appearing in the previous expressions are given by
\begin{align}
 P_1 &= m_1^4+2 m_1^2 m_2^2-2 m_1^2 m_3^2-3 m_2^4+2 m_2^2 m_3^2+m_3^4+s^2
 -2 s \left(m_1^2-m_2^2+m_3^2\right)\,, \nonumber \\
 P_2 &= m_1^8+3 m_1^6 m_2^2-7 m_1^6 m_3^2-27 m_1^4 m_2^4+18 m_1^4 m_2^2 m_3^2+9 m_1^4 m_3^4+17 m_1^2 m_2^6+29 m_1^2 m_2^4 m_3^2\nonumber \\
 &-45 m_1^2 m_2^2 m_3^4-m_1^2 m_3^6 +6 m_2^8-16 m_2^6 m_3^2+12 m_2^4 m_3^4-2 m_3^8
 -5 s^4 \nonumber \\
 &+s^3 \left(2 m_1^2-3 m_2^2+23 m_3^2\right)+s^2 \left(12 m_1^4-39 m_1^2 m_2^2-5 m_1^2 m_3^2+51 m_2^4-18 m_2^2 m_3^2-33 m_3^4\right)\nonumber \\
 &+s \left(-10 m_1^6-9 m_1^4 m_2^2+37 m_1^4 m_3^2+44 m_1^2 m_2^4-24 m_1^2 m_2^2 m_3^2-20 m_1^2 m_3^4-49 m_2^6 \right. \nonumber \\ & \left. -13 m_2^4 m_3^2+45 m_2^2 m_3^4+17 m_3^6\right)\,, \nonumber \\
 P_3 &= - m_{123} \, \mu_1 \mu_2 \mu_3 \left(m_1^4-4 m_1^2 m_2^2-2 m_1^2 m_3^2+3 m_2^4-4 m_2^2 m_3^2+m_3^4\right)\nonumber \\
 &+s^4-2 s^3 \left(2 m_1^2+3 m_2^2+2 m_3^2\right)+2 s^2 \left(3 m_1^4+3 m_1^2 m_2^2+2 m_1^2 m_3^2+6 m_2^4+3 m_2^2 m_3^2+3 m_3^4\right) \nonumber \\
 & -2 s \left(2 m_1^6-3 m_1^4 m_2^2-2 m_1^4 m_3^2-4 m_1^2 m_2^4+30 m_1^2 m_2^2 m_3^2-2 m_1^2 m_3^4+5 m_2^6\right. \nonumber \\ & \left. -4 m_2^4 m_3^2-3 m_2^2 m_3^4+2 m_3^6\right)\,,\nonumber \\
 P_4 &=m_1^6+m_1^4 m_2^2-m_1^4 m_3^2+3 m_1^2 m_2^4 -2 m_1^2 m_2^2 m_3^2-m_1^2 m_3^4
 -5 m_2^6-5 m_2^4 m_3^2+9 m_2^2 m_3^4+m_3^6
 \nonumber \\
 &-3 s^3+7 s^2 \left(m_1^2-m_2^2+m_3^2\right)
 +s \left(-5 m_1^4-2 m_1^2 m_2^2+2 m_1^2 m_3^2+15 m_2^4-10 m_2^2 m_3^2-5 m_3^4\right)\,,\nonumber \\
 P_5 &= m_1^6-7 m_1^4 m_2^2-m_1^4 m_3^2+3 m_1^2 m_2^4+14 m_1^2 m_2^2 m_3^2-m_1^2 m_3^4
 +3 m_2^6-5 m_2^4 m_3^2+m_2^2 m_3^4+m_3^6
 \nonumber \\
 &
 -s^3+s^2 \left(3 m_1^2+5 m_2^2+3 m_3^2\right)+s \left(-3 m_1^4+2 m_1^2 m_2^2-2 m_1^2 m_3^2-7 m_2^4-6 m_2^2 m_3^2-3 m_3^4\right)\,,\nonumber \\
 P_6 &= - m_{123} \, \mu_1 \mu_2 \mu_3\; \left(m_1^2-3 m_2^2-m_3^2\right) -s^3 
 + s^2 \left(3 m_1^2+7 m_2^2+m_3^2\right) \nonumber \\
 &+s \left(-3 m_1^4-2 m_1^2 m_2^2+2 m_1^2 m_3^2-3 m_2^4-14 m_2^2 m_3^2+m_3^4\right)\,,\nonumber \\
 P_7 &=- m_{123} \, \mu_1 \mu_2 \mu_3\left(m_1^2+m_2^2-m_3^2\right) +s^3 
 + s^2 \left(-m_1^2-m_2^2-3 m_3^2\right) \nonumber \\
 &+s \left(-m_1^4+10 m_1^2 m_2^2-2 m_1^2 m_3^2-m_2^4-2 m_2^2 m_3^2+3 m_3^4\right)\,.
\end{align}

Finally, we have checked all our results, both from the Feynman Parameters representation
and from the dispersion relation numerically 
against SecDec 3~\cite{Borowka:2015mxa} and pySecDec~\cite{Borowka:2017idc} and found 
perfect agreement for different values of the kinematical invariants.

\section{Conclusion}
\label{sec:conclusion}

In this paper we have presented an algorithm for computing Feynman integrals which involve 
square roots of quartic polynomials in terms of iterated integrals on the corresponding elliptic curves. 
We showed how these
iterated integrals can be reduced to a basis of elliptic integral kernels using integration-by-parts
identities. Since our kernels have only simple poles, they define a class of functions on the
elliptic curve which deserve to be called {\it elliptic multiple polylogarithms}.
The mathematical details behind the construction of these functions are
spelled out in a companion paper~\cite{longelliptic}.
Here, instead,  we have focussed
on demonstrating their use  to solve an important problem in
high energy physics which has received a lot of attention in the last decade, namely the
computation of the master integrals of the massive two-loop sunrise graph.
To demonstrate the flexibility of our approach, we studied the complete set of master integrals of
the two-loop sunrise graph in the equal and different mass case, both by direct integration
over the Feynman parameters and using a dispersion relation. 
To the best of our knowledge, this is the first time that the complete set of master integrals
of the unequal mass two-loop sunrise graph are computed explicitly in terms of elliptic 
polylogarithms.

Let us make a comment how some of the results of this paper are connected to other results in the literature.
Very recently, a paper came out~\cite{Hidding:2017jkk}, which has addressed the calculation 
of similar
integrals to the ones considered here in terms of {ad-hoc} 
elliptic generalizations of Goncharov polylogarithms. 
In contrast to ref.~\cite{Hidding:2017jkk}, we believe that our approach is more general as
it is based on a mathematically well-defined class of functions, closed under taking primitives, 
which can be proved to be substantially equivalent to the multiple
elliptic polylogarithms defined in the mathematical literature~\cite{BeilinsonLevin,LevinRacinet,BrownLevin}.
We recall here that a central result of ref.~\cite{BrownLevin} consisted in proving that
this class of functions is sufficient to integrate any rational function on a given elliptic curve.

The mathematical details of the construction of our set of functions, together with the proof of 
their equivalence
to the functions defined in ref.~\cite{BrownLevin}, are provided
in a companion paper, ref.~\cite{longelliptic}, where also many other non-trivial examples
are worked out in detail. 
Given the generality of our approach, 
we expect that our definition of elliptic polylogarithms will be applicable to a wide range of problems: from
phenomenological calculations of massive Feynman integrals in the Standard Model, to
calculations of amplitudes in $\mathcal{N}=4$ super Yang-Mills and string theory.
In particular it will be interesting to examine the recent 
representation~\cite{Bourjaily:2017bsb} of the famous ten-point
double box in $\mathcal{N}=4$ to see if it can be integrated in terms of our functions.

As a last remark, one very important aspect that we plan to study in the near future is the 
use of our functions in the context of the differential equations 
method~\cite{Kotikov:1990kg,Bern:1993kr,Remiddi:1997ny,Gehrmann:1999as}, which
has proven to be extremely powerful to compute complicated multiloop
Feynman integrals in terms of multiple polylogarithms.
In general, we expect the differential equations with respect to the kinematic invariants 
to translate in our formalism into derivatives of our functions with respect to the 
branch points that define the elliptic curve. 
A first example of this has been worked out for a special subclass of these functions
appearing in the computation of the imaginary part of the equal mass 
two-loop sunrise graph in ref.~\cite{Remiddi:2017har}.
A proper understanding of these derivatives in the general case will be essential not only to
extend the use of our functions to more complicated problems, but also to properly define
a symbol calculus for elliptic polylogarithms in a similar way as it was done
for Goncharov polylogarithms~\cite{Duhr:2011zq,Duhr:2012fh}.

\vskip1cm
\noindent\textbf{Acknowledgments: }
The authors are grateful to Nils Matthes, Erik Panzer, Brenda Penante and
Ettore Remiddi for stimulating discussions, and to the ETH Zürich and the Pauli
Center for Theoretical Studies for the organisation of the workshop ,,The
elliptic/missing Feynman integrals'', where some of the ideas presented in this
paper were first discussed. This research was supported by the the ERC grant
637019 ,,MathAm'',
the U.S. Department of Energy (DOE) under contract DE-AC02-76SF00515
and the Munich Institute for Astro- and Particle Physics
(MIAPP) of the DFG cluster of excellence ,,Origin and Structure of the
Universe''.

\appendix
\section{Integration-by-parts identities for eMPLs}
\label{app:ibp}

In this appendix we will discuss the IBP identities allowing to express a
general integral on the elliptic curve in terms of integrals over the
integration kernels provided in \eqn{eq:final_quartic}. The most general
integral on the elliptic curve has an integrand of the form
\begin{equation}
  \label{eqn:integrand}
    I(x,y) = R(x,y)\CX(x)\,,
\end{equation}
where $R(x,y)$ is a rational function, 
and $\CX(x)=Z_4(x)^\alpha \Ef{\vec n}{\vec c}{x}$ with
$\alpha$ a positive integer. The integer $\alpha +
|\vec n|$ is called the \emph{total length} of $\CX$.

The strategy is as follows: in a first step we can perform partial fractioning in $x$ to write the rational function $R(x,y)$
 in terms of members of four families of integrals. While those
families are independent with respect to partial fractioning, their members are
related through IBP identities. Correspondingly, in a second step, a set of
\textit{master integrals} must be identified within each family. 

The four families suitable for the integrand in \eqn{eqn:integrand} read
\begin{gather}
  \begin{aligned}
    \label{eqn:integralfamilies}
    A_k[\CX] &=& \int \dx x^k \CX(x),\quad\quad B_{c,k}[\CX] &=& \int \frac{\dx}{(x-c)^k}\CX(x),\\
    C_k[\CX] &=& \int \frac{\dx}{y} x^k \CX(x),\quad\quad D_{c,k}[\CX] &=& \int \frac{\dx}{y(x-c)^k}\CX(x),\\
  \end{aligned} 
\end{gather}
where $k$ is an integer.


The identification of the set of master integrals is done by recognizing
identities recursive in the variable $k$ whose recursion terminates for a
certain value of the recursion parameter. Let us show this on the example of
the family $A_k[\CX]$: they satisfy the following relation 
\begin{equation}
    \label{eqn:recursionA}
    A_k[\CX] = \frac{x^{k+1}}{k+1}\CX(x) - \frac{1}{k+1}A_{k+1}[\partial_x \CX]\,.
\end{equation}
Each time this relation is used, the value of $k$ is increased, while the
total length $\ell$ of the integrand $\CX$ in the second term is lowered through the
derivative. Reaching $\ell=0$, the integral is elementary.  However,
before getting there the recursion could reach the value $k=-1$: this integral
cannot be reduced further and is thus the master integral for the family
$A_k[\CX]$.

For the other families, similar but algebraically more involved recursion
relations exist. Here we only summarise some of the features of these recursions, and 
we refer to ref.~\cite{longelliptic} for a detailed treatment. First, the recursion for the $C$-family has 
 depth four, leading to more than one master
integral.  Second, one will have to pay attention to the case where a
shift $c$ in the integration kernels of family $D$ is a zero of the quartic
polynomial. Finally, one can identify the following set of master integrals:
\begin{align}
  \label{eqn:integralclasses}
  A_{-1} [\CX] &= \int \frac{\dx}{x} \CX(x)\,,\quad
  B_{c,1}[\CX] = \int \frac{\dx}{x-c} \CX(x)\,,\nnl
  C_{-1} [\CX] &= \int \frac{\dx}{y x} \CX(x)\,,\quad
  C_{0}  [\CX] = \int \frac{\dx}{y} \CX(x)\,,\quad
  C_{1}  [\CX] = \int \frac{x \dx}{y} \CX(x)\,,\quad
  C_{2}  [\CX] = \int \frac{x^2 \dx}{y} \CX(x)\,,\nnl
  D_{c,1}  [\CX] &= \int \frac{\dx}{y(x-c)} \CX(x)\,.
\end{align}
These integrals are in one-to-one-correspondence with the integration kernels
defined in \eqn{eq:final_quartic}: the associated integrals can then be
easily done employing \eqn{eqn:ellPoly}. 

Finally, we mention that there is one potentially problematic case: $C_2[Z_4(x)^n\textrm{E}_4]$.
This integral requires the use of special identities to rewrite powers of the
function $Z_4$. Higher powers of $Z_4$ are, however, not important for the
examples discussed in this paper.  We therefore refer the reader
to ref.~\cite{longelliptic} for a detailed discussion of this case.

\section{Integrals appearing in the second master}
\label{app:ints}
Here we list the integrals contributing to the primitive of $\CI_2$ defined in~\eqn{eqn:theIs}.
These integrals are all of the type of~\eqn{eqn:ratInt}, where $\CX(x)$ is one of the eMPLs appearing
in~\eqn{eqn:logRewritten}. Without loss of generality, we list here the integrals for $a_i=a_1$,
while they will appear summed over all four roots $a_i$ in the primitive of $\CI_2$.
\begin{align}
    \int\frac{\dx\, a_{12}a_{13}a_{14}}{y\,(x-a_1)}\Ef{-1}{0}{x}
    &=2c_4Z_4(x)\Ef{-1}{0}{x}+\mathcal{A}_0\Ef{0,-1}{0,0}{x}+\mathcal{A}_1\Ef{0}{0}{x}-(a_1-\bar{s}_1)\Ef{-1}{\infty}{x}\nnl
    &\quad-2c_4\Bigl(\Ef{-2}{0}{x}-\Ef{2}{0}{x}\Bigr)-2\Bigl(\frac{y}{x-a_1}-\frac{\bar{s}_3}{y_0}\Bigr)\Ef{-1}{0}{x}\\
 & \quad  -2\Bigl(\frac{y_0}{a_1}-\frac{\bar{s}_3}{y_0}\Bigr)\Ef{1}{0}{x}+\frac{2y_0}{a_1}\Ef{1}{a_1}{x}\,,\nnl
    \int\frac{\dx\,a_{12}a_{13}a_{14}}{y\,(x-a_1)}\Ef{-1}{1}{x}
    &=2c_4Z_4(x)\Ef{-1}{1}{x}+\mathcal{A}_0\Ef{0,-1}{0,1}{x}-2c_4\Ef{-2}{1}{x}-2\frac{y}{x-a_1}\Ef{-1}{1}{x}\nnl
    &-2\frac{y_1}{1-a_1}\Bigl(\Ef{1}{a_1}{x}-\Ef{1}{1}{x}\Bigr)\,,\\
    \int\frac{\dx\,a_{12}a_{13}a_{14}}{y\,(x-a_1)}\Ef{-1}{\infty}{x}
    &=2c_4Z_4(x)\Ef{-1}{\infty}{x}+\mathcal{A}_0\Ef{0,-1}{0,\infty}{x}-2c_4\Ef{-2}{\infty}{x}-2\frac{y}{x-a_1}\Ef{-1}{\infty}{x}\nnl
    &+2a_1\Ef{1}{a_1}{x}\,,\\
    \int\frac{\dx\,a_{12}a_{13}a_{14}}{y\,(x-a_1)}\Ef{1}{0}{x}
    &=2c_4\bigl[Z_4(x)-Z_4(0)]\Ef{1}{0}{x}+\mathcal{A}_0\Ef{0,1}{0,0}{x}-\mathcal{A}_1\Ef{0}{0}{x}-2\frac{\bar{s}_3}{y_0}\Ef{-1}{0}{x}\nnl
    &\quad-2c_4\Ef{2}{0}{x}+(a_1{-}\bar{s}_1)\Ef{-1}{\infty}{x}-2\Bigl[\frac{y}{x-a_1}+\frac{\bar{s}_3}{y_0}\Bigr]\Ef{1}{0}{x}\,.
\end{align}
Here we have defined the following abbreviations,
\begin{equation}
    \mathcal{A}_0=-\frac{3a_1^2-2a_1\,\bar{s}_1+\bar{s}_2}{3c_4}-8c_4^2\frac{\eta_1}{\omega_1}\qand
    \mathcal{A}_1=\frac{a_1\,\bar{s}_1-5\bar{s}_2}{3c_4}+\frac{8c_4\eta_1}{\omega_1}\,.
\end{equation}

\bibliographystyle{apsrev4-1} 
\bibliography{ellip} 
\end{document}